\documentclass[usenatbib]{mnras}

\pdfoutput=1

\usepackage{lineno}
\usepackage{natbib}
\usepackage{graphicx}
\usepackage{amssymb}
\usepackage{color}
\usepackage{upgreek}
\usepackage{amsmath}
\usepackage{footnote}
\usepackage{multirow}
\usepackage{tablefootnote}

\newcommand{\cursci}{Curr. Sci.}

\begin{document}

\title[IPTA Second data release]{The~International~Pulsar~Timing~Array:~Second~data~release}

\author[Perera et al.]{
\parbox{\textwidth}{
\Large
B.~B.~P.~Perera,$^{1,2}$
M.~E.~DeCesar,$^{3}$
P.~B.~Demorest,$^{4}$
M.~Kerr,$^{5}$
L.~Lentati,$^{6}$
D.~J.~Nice,$^{3}$
S.~Os{\l}owski,$^{7}$
S.~M.~Ransom,$^{8}$
M.~J.~Keith,$^{1}$
Z.~Arzoumanian,$^{9}$
M.~Bailes,$^{7,10}$
P.~T.~Baker,$^{11,12}$
C.~G.~Bassa,$^{13}$
N.~D.~R.~Bhat,$^{14}$
A.~Brazier,$^{15,16}$
M.~Burgay,$^{17}$
S.~Burke-Spolaor,$^{11,12,18}$
R.~N.~Caballero,$^{19}$
D.~J.~Champion,$^{20}$
S.~Chatterjee,$^{15}$
S.~Chen,$^{21,22,23,24}$
I.~Cognard,$^{23,21}$
J.~M.~Cordes,$^{15}$
K.~Crowter,$^{25}$
S.~Dai,$^{26}$
G.~Desvignes,$^{20,27}$
T.~Dolch,$^{28}$
R.~D.~Ferdman,$^{29}$
E.~C.~Ferrara,$^{30,31}$
E.~Fonseca,$^{32,33}$
J.~M.~Goldstein,$^{24}$
E.~Graikou,$^{20}$
L.~Guillemot,$^{23,21}$
J.~S.~Hazboun,$^{34}$
G.~Hobbs,$^{26}$
H.~Hu,$^{20}$
K.~Islo,$^{35}$
G.~H.~Janssen,$^{13,36}$
R.~Karuppusamy,$^{20}$
M.~Kramer,$^{20,1}$
M.~T.~Lam,$^{11,12}$
K.~J.~Lee,$^{37}$
K.~Liu,$^{20}$
J.~Luo,$^{38}$
A.~G.~Lyne,$^{1}$
R.~N.~Manchester,$^{39}$
J.~W.~McKee,$^{20,1}$
M.~A.~McLaughlin,$^{11,12}$
C.~M.~F.~Mingarelli,$^{40}$
A.~P.~Parthasarathy,$^{7,10}$
T.~T.~Pennucci,$^{41}$
D.~Perrodin,$^{17}$
A.~Possenti,$^{17,42}$
D.~J.~Reardon,$^{7,10}$
C.~J.~Russell,$^{43}$
S.~A.~Sanidas,$^{1}$
A.~Sesana,$^{44}$
G.~Shaifullah,$^{13}$
R.~M.~Shannon,$^{7,10}$
X.~Siemens,$^{35,45}$
J.~Simon,$^{46}$
R.~Spiewak,$^{7,10}$
I.~H.~Stairs,$^{25}$
B.~W.~Stappers,$^{1}$
J.~K.~Swiggum,$^{35}$
S.~R.~Taylor,$^{47,48}$
G.~Theureau,$^{23,21,49}$
C.~Tiburzi,$^{13}$
M.~Vallisneri,$^{46}$
A.~Vecchio,$^{24}$
J.~B.~Wang,$^{50}$
S.~B.~Zhang,$^{51,52}$
L.~Zhang,$^{53,26}$
W.~W.~Zhu,$^{54,20}$
and X.~J.~Zhu$^{55}$
}
\\
\\
Affiliations are listed at the end of the paper
\AtEndDocument{
\bigskip
\bigskip
\noindent
\emph{\small 
$^{1}$ Jodrell Bank Centre for Astrophysics, School of Physics and Astronomy, The University of Manchester, Manchester M13 9PL, UK\\
$^{2}$ Arecibo Observatory, HC3 Box 53995, Arecibo, PR 00612, USA\\
$^{3}$ Department of Physics, Lafayette College, Easton, PA 18042, USA\\
$^{4}$ National Radio Astronomy Observatory, P.O. Box O, Socorro, NM 87801, USA\\
$^{5}$ Space Science Division, Naval Research Laboratory, Washington, DC 20375-5352, USA\\
$^{6}$ Astrophysics Group, Cavendish Laboratory, JJ Thomson Avenue, Cambridge CB3 0HE, UK\\
$^{7}$ Centre for Astrophysics and Supercomputing, Swinburne University of Technology, Post Box 218 Hawthorn, VIC 3122, Australia\\
$^{8}$ National Radio Astronomy Observatory, 520 Edgemont Road, Charlottesville, VA 22903, USA\\
$^{9}$ Code 662, X-ray Astrophysics Laboratory, NASA Goddard Space Flight Center, Greenbelt, MD 20770 USA\\
$^{10}$ ARC Centre of Excellence for Gravitational Wave Discovery (OzGrav)\\
$^{11}$ Department of Physics and Astronomy, West Virginia University, P.O. Box 6315, Morgantown, WV 26506, USA\\
$^{12}$ Center for Gravitational Waves and Cosmology, West Virginia University, Chestnut Ridge Research Building, Morgantown, WV 26505, USA\\
$^{13}$ ASTRON, the Netherlands Institute for Radio Astronomy, Oude Hoogeveensedijk 4, Dwingeloo 7991 PD, the Netherlands\\
$^{14}$ International Centre for Radio Astronomy Research, Curtin University, Bentley, WA 6102, Australia\\
$^{15}$ Department of Astronomy and Cornell Center for Astrophysics and Planetary Science, Cornell University, Ithaca, NY 14853, USA\\
$^{16}$ Cornell Center for Advanced Computing, Cornell University, Ithaca, NY 14853, USA\\
$^{17}$ INAF-Osservatorio Astronomico di Cagliari, via della Scienza 5, I-09047 Selargius\\
$^{18}$ Canadian Institute for Advanced Research, CIFAR Azrieli Global Scholar, MaRS Centre West Tower, 661 University Ave. Suite 505, Toronto ON M5G 1M1, Canada\\
$^{19}$ Kavli Institute for Astronomy and Astrophysics, Peking University, Beijing 100871, China\\
$^{20}$ Max-Planck-Institut f$\ddot u$r Radioastronomie, Auf dem H$\ddot u$gel 69, D-53121 Bonn, Germany\\
$^{21}$ Station de radioastronomie de Nan{\c c}ay, Observatoire de Paris, PSL Research University, CNRS/INSU F-18330 Nan{\c c}ay, France\\
$^{22}$ FEMTO-ST Institut de recherche, Department of Time and Frequency, UBFC and CNRS, ENSMM, F-25030 Besan\c{c}on, France\\
$^{23}$ Laboratoire de Physique et Chimie de l'Environnement et de l'Espace LPC2E CNRS-Universit{\'e} d'Orl{\'e}ans, F-45071 Orl{\'e}ans, France\\
$^{24}$ School of Physics and Astronomy and Institute for Gravitational Wave Astronomy, University of Birmingham, Birmingham, B15 2TT, UK\\
$^{25}$ Department of Physics and Astronomy, University of British Columbia, 6224 Agricultural Road, Vancouver, BC, V6T 1Z1, Canada\\
$^{26}$ CSIRO Astronomy \& Space Science, Australia Telescope National Facility, P.O. Box 76, Epping, NSW 1710, Australia\\
$^{27}$ LESIA, Observatoire de Paris, Universit\'e PSL, CNRS, Sorbonne Universit\'e, University Paris Diderot, Sorbonne Paris Cit\'e, 5 place Jules Janssen, 92195 Meudon, France\\
$^{28}$ Department of Physics, Hillsdale College, 33 E. College Street, Hillsdale, Michigan 49242, USA\\
$^{29}$ School of Chemistry, University of East Anglia, Norwich Research Park, Norwich NR4 7TJ, UK\\
$^{30}$ Department of Astronomy, University of Maryland, College Park, MD 20742 USA\\
$^{31}$ NASA Goddard Space Flight Center, Greenbelt, MD 20770 USA\\
$^{32}$ Department of Physics, McGill University, 3600 rue University, Montr\'eal, QC H3A 2T8, Canada\\
$^{33}$ McGill Space Institute, McGill University, 3550 rue University, Montr\'eal, QC H3A 2A7, Canada\\
$^{34}$ Physical Sciences Division, University of Washington Bothell, Bothell, WA 98011, USA\\
$^{35}$ Center for Gravitation, Cosmology, and Astrophysics, Department of Physics, University of Wisconsin--Milwaukee, PO Box 413, Milwaukee, WI 53201, USA\\
$^{36}$ Department of Astrophysics/IMAPP, Radboud University, P.O. Box 9010, 6500 GL Nijmegen, The Netherlands\\
$^{37}$ Kavli institute for astronomy and astrophysics, Peking University, Beijing 100871, China\\
$^{38}$ Canadian Institute for Theoretical Astrophysics, University of Toronto, 60 Saint George Street, 14th floor, Toronto, ON, M5S 3H8, Canada\\
$^{39}$ CSIRO Astronomy and Space Science, Australia Telescope National Facility, PO Box 76, Epping 1710 NSW, Australia\\
$^{40}$ Center for Computational Astrophysics, Flatiron Institute, 162 Fifth Ave, New York, NY 10010, USA\\
$^{41}$ Hungarian Academy of Sciences MTA-ELTE ``Extragalatic Astrophysics'' Research Group, Institute of Physics, E\"{o}tv\"{o}s Lor\'{a}nd University, P\'{a}zm\'{a}ny P. s. 1/A, Budapest 1117, Hungary\\
$^{42}$ Università di Cagliari, Dept of Physics, S.P. Monserrato-Sestu Km 0,700, I-09042 Monserrato\\
$^{43}$ CSIRO Scientific Computing Services, Australian Technology Park, Locked Bag 9013, Alexandria, NSW 1435, Australia\\
$^{44}$ Universit\'a di Milano Bicocca, Dipartimento di Fisica, Piazza della Scienza 3, 20126, Milan, Italy\\
$^{45}$ Department of Physics, Oregon State University, Corvallis, OR 97331, USA\\
$^{46}$ Jet Propulsion Laboratory, California Institute of Technology, Pasadena CA 91109, USA\\
$^{47}$ TAPIR Group, California Institute of Technology, 1200 East California Boulevard, Pasadena 91125, California, USA\\
$^{48}$ Department of Physics \& Astronomy, Vanderbilt University, 2301 Vanderbilt Place, Nashville, TN 37235, USA\\
$^{49}$ Laboratoire Univers et Th[\'e]ories LUTh, Observatoire de Paris, PSL Research University, CNRS/INSU, Universit{\'e} Paris Diderot, 5 place Jules Janssen, 92190 Meudon, France\\
$^{50}$ Key Laboratory of Radio Astronomy, Chinese Academy of Science, 150 Science 1-Street, Urumqi, Xinjiang 830011, China\\
$^{51}$ Purple Mountain Observatory, Chinese Academy of Sciences, Nanjing 210008, China\\
$^{52}$ University of Chinese Academy of Sciences, Beijing 100049, China\\
$^{53}$ National Astronomical Observatories, Chinese Academy of Sciences, A20 Datun Road, Chaoyang District, Beijing 100101, China\\
$^{54}$ CAS Key Laboratory of FAST, NAOC, Chinese Academy of Sciences, Beijing 100101, China\\
$^{55}$ OzGrav-Monash, School of Physics and Astronomy, Monash University, Clayton, VIC 3800, Australia\\
}
}
}

\maketitle

\begin{abstract}
In this paper, we describe the International Pulsar Timing Array second data release, which includes recent pulsar timing data obtained by three regional consortia: the European Pulsar Timing Array, the North American Nanohertz Observatory for Gravitational Waves, and the Parkes Pulsar Timing Array. We analyse and where possible combine high-precision timing data for 65 millisecond pulsars which are regularly observed by these groups. A basic noise analysis, including the processes which are both correlated and uncorrelated in time, provides noise models and timing ephemerides for the pulsars. We find that the timing precisions of pulsars are generally improved compared to the previous data release, mainly due to the addition of new data in the combination. The main purpose of this work is to create the most up-to-date IPTA data release. These data are publicly available for searches for low-frequency gravitational waves and other pulsar science.
\end{abstract}


\section{Introduction}
 
Pulsar timing observations are sensitive to correlated signals at low frequencies, from nHz to $\upmu$Hz, such as those caused by gravitational waves (GWs) produced from inspiraling supermassive black hole binaries (SMBHBs). Millisecond pulsars (MSPs) have been identified as ideal tools for searching for GWs due to their excellent rotational stability \citep[see][]{det79,hd83,jhlm05}. They are old neutron stars that are spun up to spin periods of $\lesssim$20~ms during an accretion phase \citep[``recycling'':][]{acrs82,rs82}. High-precision timing measurements of many MSPs with sub-microsecond precision---a Pulsar Timing Array (PTA)---collected over long time spans offer a unique and powerful probe of low-frequency GWs \citep{dcl+16,abb+15,rhc+16}. The International Pulsar Timing Array\footnote{\url{http://ipta4gw.org}} (IPTA) seeks to further improve the sensitivity of PTAs by combining the data from three individual PTAs, namely the European Pulsar Timing Array \citep[EPTA;][]{dcl+16}, the North American Nanohertz Observatory for Gravitational Waves \citep[NANOGrav:][]{abb+18}, and the Parkes Pulsar Timing Array \citep[PPTA;][]{rhc+16}. The combination of all the data from the individual PTAs under the auspices of the IPTA should reduce the time to the detection of GWs: the GWs from the cosmic merger history of SMBHBs should create a GW background which may be detectable in the next five years \citep{sej+13,rsg15,kbh+17,tve+16}, and GWs from individual SMBHBs in the next ten years \citep{rsg15,mls+17,kbh+18}.

The first IPTA data release \citep[IPTA~dr1 --][]{vlh+16} reported a combination of timing data of 49 MSPs observed by individual PTAs. The data lengths of these pulsars ranged between 4.5 - 27 years, depending on when the source was included in the timing campaign. The data release included the timing data from the EPTA until February 2013, NANOGrav until October 2009, and the PPTA until October 2013. Recently, the EPTA \citep{dcl+16}, NANOGrav \citep{abb+15}, and the PPTA \citep{rhc+16} reported new data releases. Here we report the creation of the IPTA second data release (IPTA dr2) and make it available for GW search experiments and other related science. We note that the recently-released NANOGrav 11-year data set \citep{abb+18} and the new PPTA dr2 (Kerr et al. in preparation) will be included in future IPTA data releases.

The timing data released by individual PTAs has been used to search for GWs and place upper limits on their strain amplitudes \citep[see][]{yhj+10,zhw+14,abb+14,bps+16,ltm+15,srl+15,abb+16, abb+18,psb+18,aab+18}. The IPTA dr1 has also been used in GW search experiments and has placed limits on the stochastic GW background \citep{vlh+16}. Furthermore, with a better sky-coverage, \citet{gvs+18} addressed the importance of the IPTA data set in localising resolvable GW sources and reported that the results are superior to what is achieved by individual PTAs. \cite{mab+19} showed how one can combine IPTA dr1 with \cite{gaiadr2} data to improve binary pulsar distance estimates, which can in turn be used to improve PTA sensitivity to individual SMBHB systems \citep{cc10,lwk+11,ellis13,teg14,zwx+16} and eventually measure their spin \citep{sv10,mgs+12}. 
\citet{lsc+16} showed the importance of the IPTA dr1 by studying the noise processes of pulsars to improve their timing stabilities. Therefore, a more up-to-date IPTA data combination is crucial to improve the timing precision and thus, the sensitivity of pulsars to GWs, leading towards a detection in the near future.  In addition to the search for GWs, the IPTA data set has been used in other areas of astrophysics. For example, \citet{cgl+18} utilised the IPTA dr1 to study the solar system and provided improved PTA mass estimates for planetary systems, the first PTA-based estimates of asteroid-belt object masses, such as the dwarf planet Ceres, and provided generic mass limits for unknown objects in orbits in the solar system, including theoretical objects such as dark matter clumps.

\citet{vlh+16} described the pulsar timing, procedure for creating and combining IPTA data sets, and the usage of the IPTA data  comprehensively. The process of the new data combination here in IPTA dr2 is broadly similar to that of the IPTA dr1 and thus we only briefly overview the combination procedure in this paper, and refer the reader to \citet{vlh+16} for additional details.
The paper is organised as follows: we first describe the constituent PTA data sets used in this combination in \S~\ref{data}. The data combination procedure is briefly described in \S~\ref{procedure} and the final data products are presented in \S~\ref{results}. We discuss our results and compare with the results of IPTA dr1 in \S~\ref{dis}. Finally in \S~\ref{future}, we discuss the future projects that will be carried out using this new IPTA data release.

\section{Data sets}
\label{data}
To produce the IPTA dr2, we combined published data from recent individual PTA data releases, along with a selection of additional data sets that were either used in the IPTA dr1 or published in other studies. Detailed descriptions of each of these data sets are given below. 

\smallskip

\noindent
{\bf EPTA data set:} We include the most recent EPTA data release 1.0 \citep{dcl+16} in the IPTA dr2. This data set includes high-precision timing observations from 42 MSPs obtained with the Effelsberg Radio Telescope (EFF) in Germany, the Lovell Radio Telescope at the Jodrell Bank Observatory (JBO) in the UK, the Nan\c{c}ay Radio Telescope (NRT) in France, and the Westerbork Synthesis Radio Telescope (WSRT) in the Netherlands. The data set spans timing baselines of 7 to 18 years, covering from October 1996 to January 2015. In addition, we note that the data set of PSR~J1939$+$2134 includes very early NRT observations that started in March 1990. Each observation is averaged both in time and frequency, across the bandwidth, and provides a single time-of-arrival (ToA). Observation information is given in Table~\ref{info} and additional details can be found in \citet{dcl+16}.

\smallskip

\noindent
{\bf NANOGrav data set:} We include the NANOGrav 9 year data set \citep{abb+15} in this data combination. This includes high-precision timing observations obtained from 37 MSPs, with timing baselines between 0.6--9.2\,years from July 2004 to March 2013. We also include the long-term NANOGrav timing data of PSR J1713$+$0747 reported in \citet{zsd+15}, and the data of PSRs J1857+0943 and J1939+2134 from November 1984 through December 1992 reported in \citet{ktr94}. All of these observations were obtained using the Robert C.~Byrd Green Bank Telescope (GBT) and the Arecibo Observatory (AO) in the USA. We note that the ToAs in the NANOGrav 9 year data set are obtained by first averaging the observations in time as given in \citet{abb+15}, and in frequency such that the data maintain a frequency resolution (i.e. sub-band information) ranging from 1.5 to 12.5~MHz depending on the combination of receiver and backend\footnote{The frequency channel bandwidths in the NANOGrav 9-year data set are: ASP/GASP: 4~MHz at all frequencies; PUPPI/GUPPI: 1.6~MHz at below 500~MHz; 3.1~MHz between 500 and 1000~MHz; and 12.5~MHz above 1000~MHz.}. Each frequency channel yields a single ToA.
The observations in \citet{zsd+15} are partially averaged in time and frequency, resulting in multiple ToAs for a given observation epoch. The observations in \citet{ktr94} are fully averaged in time and frequency, leading to one ToA for each receiver and data acquisition system at each epoch.

\smallskip

\noindent
{\bf PPTA data set:} We include the PPTA first data release \citep{mhb+13} and its extended version \citep{rhc+16} in this IPTA data combination.
This PPTA data set includes high-precision timing observations obtained from 20 MSPs with an observation time baseline of approximately six years. Additional ``legacy'' L-band (i.e. 1400~MHz) observations acquired between 1994 and 2005, for which the raw data are no longer available, are also included in the combination. 
Finally, we include more recent PPTA observations reported in \citet{srl+15} for the high-precision PSRs J0437$-$4715, J1744$-$1134, J1713$+$0747, and J1909$-$3744. All the PPTA observations are obtained using the Parkes Radio Telescope in Australia and a range of receivers and pulsar timing backends.
Although the ToA coverage is nearly identical to the data sets indicated above, the raw data from 2005 onwards have been reprocessed using a pipeline developed for new PPTA data releases (Kerr et al. in preparation).  
In general, PPTA data is divided into four bands with wavelengths of roughly 10, 20, 40, and 50\,cm.  An analytic template for the pulse profile for each instrument and band is produced, and the unknown phase offset between these templates is measured from the data as a free parameter in the timing model.  Instrumental offsets (``JUMPs'') were obtained using a modulated PIN diode as described in \citet{mhb+13}.  Similar to the EPTA data set, each observation of the PPTA data set is averaged in time and frequency, resulting in a single ToA for each radio receiver at each epoch.

\begin{figure}
\includegraphics[width=8.5cm]{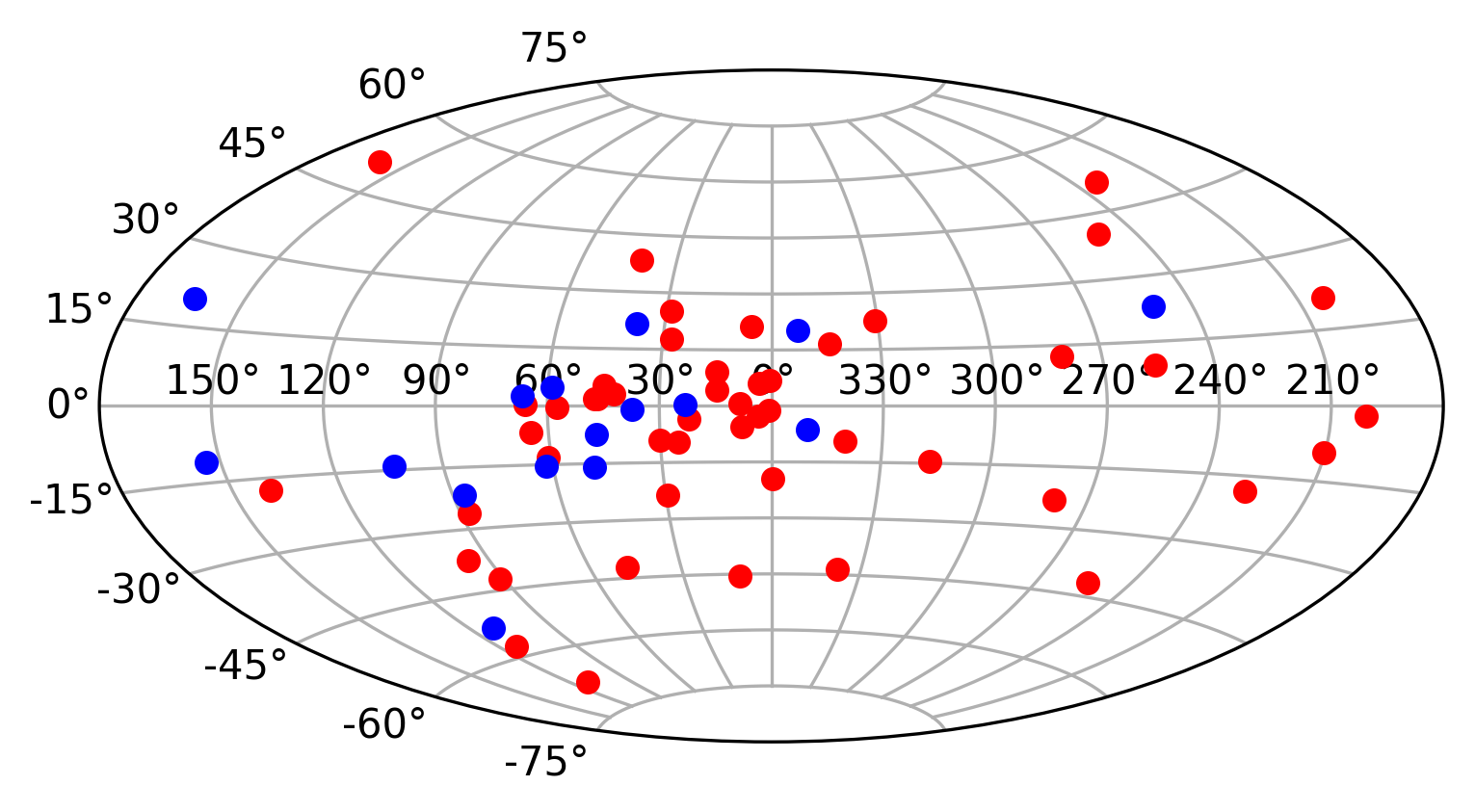}
\caption{
The Galactic distribution of 65 pulsars in the data release, including 49 pulsars from the IPTA dr1 ({\it red} dots) and 16 new pulsars ({\it blue} dots) that were not presented in the IPTA dr1. Galactic latitude is on the vertical axis in degrees, and Galactic longitude is on the horizontal axis in degrees, increasing leftward with the Galactic centre at the origin. Note that many new pulsars included in this data release fill some gaps in the IPTA dr1 pulsar distribution.
}
\label{aitoff}
\end{figure}

\begin{figure*}
\includegraphics[width=18cm]{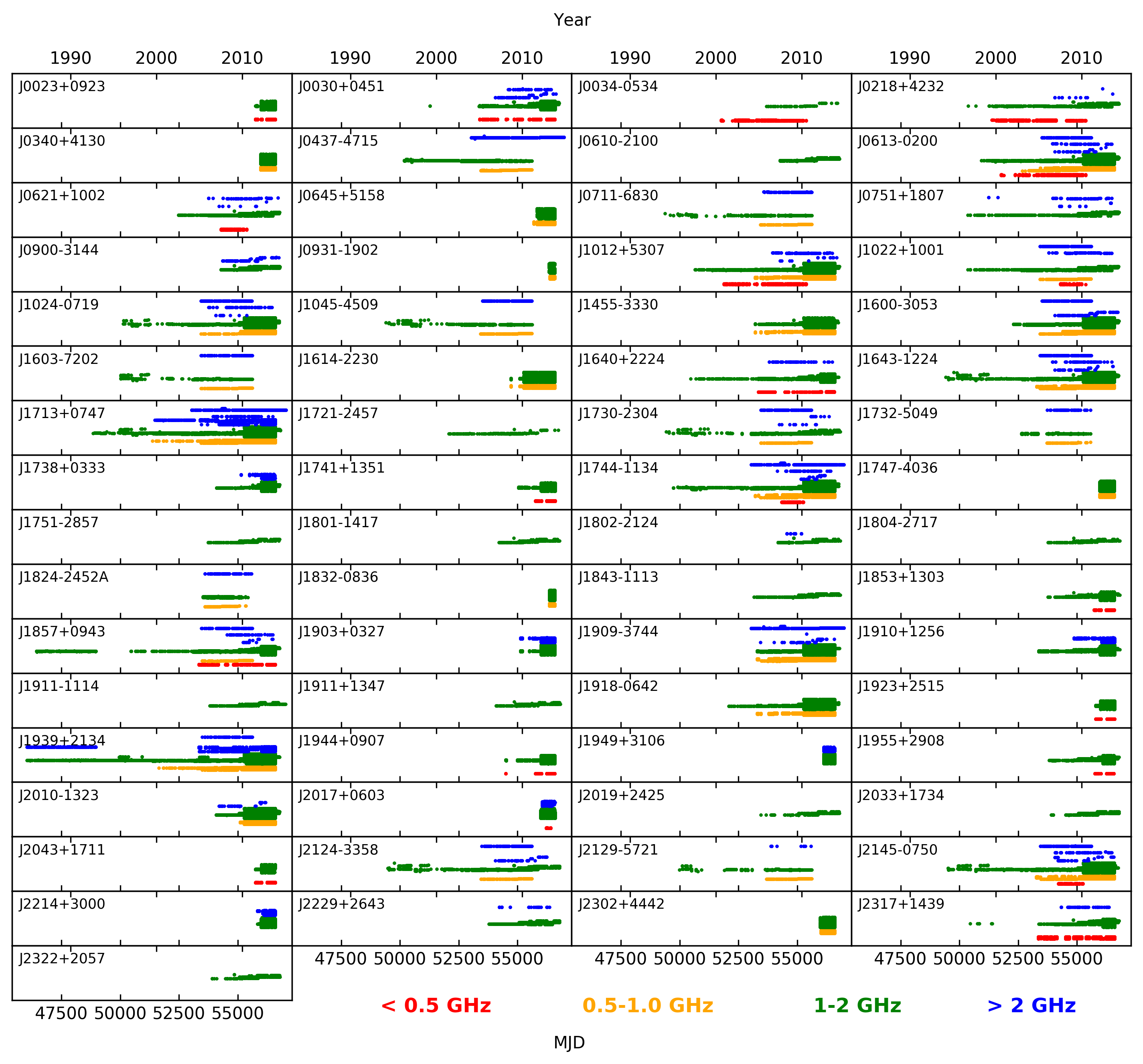}
\caption{
The frequency coverage and the time baseline of the observations used in IPTA dr2 for each pulsar. Note that all pulsars are observed at L-band ($\sim$1400~MHz). The y-axis of each panel represents a frequency range of approximately 0--4~GHz in linear scale. The frequency of each ToA is plotted, so that approximately the entire bandwidth is shown for NANOGrav observations because of their available sub-band ToAs.  
}
\label{psrs_freqs}
\end{figure*}

Combining all the above mentioned data sets, the new IPTA data release comprises 65 pulsars in total, adding 16 new pulsars compared to IPTA dr1. All of these new pulsars are observed and included by the NANOGrav timing campaign. By comparing positions, it is evident that these new pulsars improved the IPTA pulsar distribution in the Galaxy, providing a better sky coverage compared to the previous data release (see Figure~\ref{aitoff}). A summary of the data sets used in this data release is given in Table~\ref{info} and the basic parameters of these MSPs are given in Table~\ref{psrs_par}. Figure~\ref{psrs_freqs} shows the frequency coverages and the time baselines of these data sets in the data combination.

\begin{table*}
\begin{center}
\caption{
The observation information in PTA data releases. Note that, as in IPTA dr1, the GBT and AO observations of PSR J1713$+$0747 reported in \citet{zsd+15} and early AO observations of PSRs J1857$+$0943 and J1939$+$2134 reported in \citet{ktr94} are included in the data combination. 
}
\label{info}
\begin{tabular}{llcccl}
\hline
\multicolumn{1}{l}{PTA} &
\multicolumn{1}{l}{Telescope} &
\multicolumn{1}{c}{Typical} &
\multicolumn{1}{c}{No. of} &
\multicolumn{1}{c}{Observing} &
\multicolumn{1}{c}{Data span} \\
\multicolumn{1}{c}{ } &
\multicolumn{1}{c}{} &
\multicolumn{1}{c}{cadence} &
\multicolumn{1}{c}{pulsars} &
\multicolumn{1}{c}{Frequencies} &
\multicolumn{1}{c}{(MJD/Gregorian)} \\
\multicolumn{1}{c}{ } &
\multicolumn{1}{c}{ } &
\multicolumn{1}{c}{(weeks) } &
\multicolumn{1}{c}{ } &
\multicolumn{1}{c}{ (GHz) } &
\multicolumn{1}{c}{ Earliest$-$Latest } \\
\hline
EPTA & EFF & 4 & 18 & 1.4, 2.6 & 50360 (1996 Oct 04) $-$ 56797 (2014 May 20) \\
 & JBO & 3 & 35 & 1.4 & 54844 (2009 Jan 13) $-$ 57028 (2015 Jan 06) \\
 & NRT & 2 & 42 & 1.4, 2.1 & 47958 (1990 Mar 08) $-$ 56810 (2014 Jun 02) \\
 & WSRT & 4 & 19 & 0.3, 1.4, 2.2 & 51386 (1999 Jul 27) $-$ 55375 (2010 Jun 28) \smallskip \\
NANOGrav & GBT & 4 & 20 & 0.8, 1.4 & 53216 (2004 Jul 30) $-$ 56598 (2013 Nov 02) \\
 & AO & 4 & 19 & 0.3, 0.4, 1.4, 2.3 & 53343 (2004 Dec 04) $-$ 56599 (2013 Nov 03) \\
\citet{zsd+15} & GBT and AO & 2 & 1 & 0.8, 1.4, 2.3 & 48850 (1992 Aug 16) $-$ 56598 (2013 Nov 02) \\
\citet{ktr94} & AO & 2 & 2 & 1.4, 2.3 & 46436 (1986 Jan 06) $-$ 48973 (1992 Dec 17) \smallskip \\
PPTA & PKS & 2 & 20 & 0.6, 1.4, 3.1 & 49373 (1994 Jan 21) $-$ 57051 (2015 Jan 29) \\
\hline
\end{tabular}
\end{center}
\end{table*}

\begin{table*}
\begin{center}
\caption{
The basic properties of the pulsars in the IPTA second data release. The sources in this new data release that were not in IPTA~dr1 are marked with  `$^*$'.
The flux density of the pulsar at 1.4~GHz is quoted in the fourth column. The distance to the pulsar is given in fifth column, using the electron density model YMW16 \citep{ymw17} based on the timing measured DM value, or compiled by the model given in \citet{vwc+12}, either using the updated parallax measurement from this paper (denoted by $\dagger$ -- see Appendix~\ref{app}) or from previously published parallax measurements$^a$ (denoted by $\ddag$).
The uncertainty of the DM-derived distance is estimated considering a typical 20 per cent error of the electron density model. The next columns indicate with an `X'  whether the pulsar is observed by that particular PTA. The ninth column presents if the DM model given in \citet{kcs+13} is included (Y) or excluded (N) in the timing solution of the pulsar according to VersionA (see \S~\ref{verA}). In the tenth column, we quote the weighted root mean square of the timing residuals $\sigma_w$, after subtracting out the timing model and the maximum likelihood time-correlated signals reported in VersionB (see \S~\ref{verB}). The eleventh column gives the time span of the data set. }
\label{psrs_par}
\begin{tabular}{lrrrlccccccl}
\hline
\multicolumn{1}{l}{PSR} &
\multicolumn{1}{c}{Pulse} &
\multicolumn{1}{c}{DM} &
\multicolumn{1}{r}{S$_{\rm 1.4}$} &
\multicolumn{1}{c}{Distance} &
\multicolumn{1}{c}{\parbox[t]{2mm}{\multirow{5}{*}{\rotatebox[origin=c]{90}{EPTA}}}} &
\multicolumn{1}{c}{\parbox[t]{2mm}{\multirow{5}{*}{\rotatebox[origin=c]{90}{NANOGrav}}}} &
\multicolumn{1}{c}{\parbox[t]{2mm}{\multirow{5}{*}{\rotatebox[origin=c]{90}{PPTA}}}} &
\multicolumn{1}{c}{DM$_k$} &
\multicolumn{1}{c}{$\sigma_{w}$} &
\multicolumn{1}{c}{Span} &
\multicolumn{1}{c}{References } \\
\multicolumn{1}{c}{} &
\multicolumn{1}{c}{period} &
\multicolumn{1}{c}{(cm$^{-3}$~pc)} &
\multicolumn{1}{r}{(mJy)} &
\multicolumn{1}{c}{(kpc)} &
\multicolumn{1}{c}{} &
\multicolumn{1}{c}{} &
\multicolumn{1}{c}{} &
\multicolumn{1}{c}{} &
\multicolumn{1}{c}{($\mu$s)} &
\multicolumn{1}{c}{(year)} &
\multicolumn{1}{c}{}  \\
\multicolumn{1}{c}{} &
\multicolumn{1}{c}{(ms)} &
\multicolumn{1}{c}{ } &
\multicolumn{1}{r}{} &
\multicolumn{1}{c}{ } &
\multicolumn{1}{c}{} &
\multicolumn{1}{c}{} &
\multicolumn{1}{c}{} &
\multicolumn{1}{c}{} &
\multicolumn{1}{c}{} &
\multicolumn{1}{c}{} &
\multicolumn{1}{c}{}  \\\\\\
\hline
J0023$+$0923$^*$ & 3.050  & 14.33 & 0.5 & $1.2\pm0.2$ & & X & & Y & 1.34& 2.3 &1, 2, 3 \\
J0030$+$0451 & 4.865  & 4.33 & 0.6 & $0.34\pm0.01\dagger$  & X & X & & Y & 1.48 & 15.1 & 4, 2, 3 \\
J0034$-$0534 & 1.877 & 13.77 & 0.6 & $1.03\pm0.3$ & X & & & N & 4.19 & 13.5 & 5, 6, 7 \\
J0218$+$4232 & 2.323 & 61.25 & 0.9 & $3.7_{-0.8}^{+1.1}\ddag$ & X & & & Y & 7.01 & 17.6 & 8, 9, 10, 11, 12  \\
J0340$+$4130$^*$ & 3.299 & 49.58 & 0.3 & $1.6\pm0.3$ & & X & & Y & 5.16 & 1.7 & 1, 2, 3 \\
J0437$-$4715  & 5.757 & 2.64 & 149.0 & $0.156\pm0.001\ddag$ & & & X & Y & 0.11 & 18.6 & 13, 14, 15, 16, 17 \\
J0610$-$2100 & 3.861 & 60.67 & 0.4 & $3.3\pm0.7$ & X & & & N & 4.88 & 6.9 & 18 \\
J0613$-$0200 & 3.062 & 38.78 & 2.3 & $1.11\pm0.05\dagger$ & X & X & X & Y & 1.14 & 16.0 & 19, 2, 20, 21 \\
J0621$+$1002 & 28.854 & 36.47 & 1.9 & $0.4\pm0.1$ & X & & & Y & 6.57 & 11.8 & 22, 9, 11 \\
J0645$+$5158$^*$ & 8.853 & 18.25 & 0.3 & $0.7\pm0.1$ & & X & & Y & 0.57 & 2.4 & 23, 2, 3 \\
J0711$-$6830 & 5.491 & 18.41 & 3.2 & $0.11\pm0.02$ & & & X & Y & 1.44 & 17.1 & 24, 14, 20, 16  \\
J0751$+$1807 & 3.479 & 30.25 & 3.2 & $1.4_{-0.3}^{+0.4}\ddag$ & X & & & N & 3.00 & 17.6 & 25, 9, 11, 26 \\
J0900$-$3144 & 11.110 & 75.71 & 3.8 & $0.4\pm0.1$ & X & & & N & 3.21 & 6.9 & 18, 9, 16  \\
J0931$-$1902$^*$ & 4.638 & 41.49 & 0.4 & $3.7\pm0.7$ & & X & & N & 3.69 & 0.6 & 27, 2, 3 \\
J1012$+$5307 & 5.256 & 9.02 & 3.2 & $0.8_{-0.1}^{+0.2}\ddag$ & X & X & & Y & 1.91& 16.8 & 28, 29, 3 \\
J1022$+$1001 & 16.453 & 10.25 & 6.1 & $0.72\pm0.02\ddag$ & X & & X & Y & 1.97 & 17.5 & 22, 14, 20, 21, 30  \\
J1024$-$0719 & 5.162  & 6.49 & 1.5 & $1.2_{-0.1}^{+0.2}\dagger$ & X & X & X & Y & 1.71 & 18.2 & 24, 2, 16  \\
J1045$-$4509 & 7.474 & 58.14 & 2.7 & $0.5_{-0.3}^{+1.3}\ddag$ & & & X & Y & 3.19 & 17.0 & 5, 14, 16, 31  \\
J1455$-$3330 & 7.987 & 13.57 & 1.2 & $1.0_{-0.2}^{+0.3}\ddag$ & X & X & & Y & 4.12 & 9.7 & 19, 2, 7, 26 \\
J1600$-$3053 & 3.598 & 52.32 & 2.5 & $2.0_{-0.2}^{+0.3}\dagger$ & X & X & X & Y & 0.92 & 12.3 & 32, 2, 21 \\
J1603$-$7202 & 14.842 & 38.05 & 3.1 & $1.1\pm0.2$ & & & X & Y & 1.58 & 15.3 & 33, 14, 20, 21 \\
J1614$-$2230$^*$ & 3.151 & 34.49 & 0.7 & $0.69_{-0.04}^{+0.05}\dagger$ & & X & & Y & 1.38 & 5.1 & 34, 2, 3 \\
J1640$+$2224 & 3.163 & 18.42 & 2.0 & $1.5\pm0.3$ & X & X & & Y & 0.77 & 17.2 & 2, 11 \\
J1643$-$1224 & 4.622 & 62.41 & 4.8 & $1.1_{-0.3}^{+0.6}\dagger$ & X & X & X & Y & 2.55 & 20.1 & 19, 14, 21 \\
J1713$+$0747 & 4.570 & 15.97 & 10.2 & $1.20\pm0.03\dagger$ & X & X & X & Y & 0.21 & 22.5 & 35, 2, 21 \\
J1721$-$2457 & 3.497  & 47.76 & 0.6 & $1.4\pm0.3$ & X & & & N & 12.21 & 12.8 & 36, 37 \\
J1730$-$2304 & 8.123 & 9.62 & 3.9 & $0.60_{-0.07}^{+0.09}\dagger$ & X & & X & Y & 1.57 & 20.3 & 19, 14, 16 \\
J1732$-$5049 & 5.313 & 56.82 & 1.3 & $1.87\pm0.4$ & & & X & Y & 2.72 & 8.0 & 36, 14, 21 \\
J1738$+$0333 & 5.850 & 33.77 & 0.7 & $1.5\pm0.1\ddag$ & X & X & & Y & 1.38 & 7.3 & 38, 39, 3  \\
J1741$+$1351$^*$ & 3.747 & 24.20 & 0.9 & $1.4\pm0.3$ & & X & & Y & 0.46 & 4.2 & 32, 2 \\
J1744$-$1134 & 4.075 & 3.137 & 3.1 & $0.410\pm0.008\dagger$ & X & X & X & Y & 0.73 & 19.9 & 24, 14, 16 \\
J1747$-$4036$^*$ & 1.646 & 152.98 & 0.9 & $7.1\pm1.4$ & & X & & Y & 4.79 & 1.7 & 40, 2, 41 \\
J1751$-$2857 & 3.915 & 42.84 & 0.1 & $1.1\pm0.2$ & X & & & N & 2.85 & 8.3 & 42, 9 \\
J1801$-$1417 & 3.625 & 57.26 & 0.2 & $1.1\pm0.2$ & X & & & N & 2.76 & 7.0 & 43, 9, 44 \\
J1802$-$2124 & 12.648 & 149.63 & 0.8 & $3.0\pm0.6$ & X & & & N & 2.76 & 7.2 & 43, 45 \\
J1804$-$2717 & 9.343 & 24.67 & 0.4 & $0.8\pm0.2$& X & & & N & 3.72 & 8.4 & 33, 9, 10, 11 \\
\hline
\end{tabular}
\begin{tabular}{l}
$^a$\url{http://hosting.astro.cornell.edu/research/parallax/}\\
{\it References}: (1)~\citet{hrm+11}, (2)~\citet{abb+18}, (3)~\citet{lmj+16},
(4)~\citet{lzb+00}, \\
(5)~\citet{bhl+94}, (6)~\citet{aaa+10e}, (7)~\citet{tbms98}, 
(8)~\citet{nbf+95}, (9)~\citet{dcl+16}, \\
(10)~\citet{hlk+04}, (11)~\citet{kxl+98}, (12)~\citet{dyc+14},
(13)~\citet{jlh+93}, (14)~\citet{rhc+16}, \\
(15)~\citet{vbv+08}, (16)~\citet{jvk+18}, (17)~\citet{dvtb08},d
(18) ~\citet{bjd+06}, 
(19)~\citet{lnl+95}, \\
(20)~\citet{hbo06}, (21)~\citet{mhb+13}, 
(22)~\citet{cnst96}, 
(23)~\citet{slr+14},
(24)~\citet{bjb+97},\\
(25)~\citet{lzc95}, (26)~\citet{gsl+16}, 
(27)~\citet{abb+15}, 
(28)~\citet{nll+95}, \\(29)~\citet{lwj+09}, 
(30)~\citet{dgb+19}, 
(31)~\citet{vbc+09}, 
(32)~\citet{jbo+07},
(33)~\citet{llb+96},\\
(34)~\citet{crh+06a}
(35)~\citet{fwc93},
(36)~\citet{eb01b}, (37)~\citet{jsb+10},
(38)~\citet{jac05}, \\(39)~\citet{fwe+12}, 
(40)~\citet{kcj+12}, (41)~\citet{ckr+15}, 
(42)~\citet{sfl+05},
(43)~\citet{fsk+04}, \\(44)~\citet{lfl+06}, 
(45)~\citet{fsk+10},
(46)~\citet{lbm+87},
(47)~\citet{bbb+13}, 
(48)~\citet{hfs+04},\\
(49)~\citet{srs+86},
(50)~\citet{crl+08}, (51)~\citet{fbw+11},
(52)~\citet{jbv+03}, 
(53)~\citet{lbr+13}, \\
(54)~\citet{bkh+82},
(55)~\citet{clm+05},
(56)~\citet{dfc+12},
(57)~\citet{bbf83}, 
(58)~\citet{cgj+11},\\
(59)~\citet{ntf93}, (60)~\citet{nss01},
(61)~\citet{rtj+96}, 
(62)~\citet{rrc+11}, 
(63)~\citet{cam95a}, 
(64)~\citet{cnt93}
\end{tabular}
\end{center}
\end{table*}

\begin{table*}
\begin{center}
\contcaption{}
\begin{tabular}{lrrrlccccccl}
\hline
\multicolumn{1}{l}{PSR} &
\multicolumn{1}{c}{Pulse} &
\multicolumn{1}{c}{DM} &
\multicolumn{1}{r}{F$_{\rm 1.4}$} &
\multicolumn{1}{c}{Distance} &
\multicolumn{1}{c}{\parbox[t]{2mm}{\multirow{5}{*}{\rotatebox[origin=c]{90}{EPTA}}}} &
\multicolumn{1}{c}{\parbox[t]{2mm}{\multirow{5}{*}{\rotatebox[origin=c]{90}{NANOGrav}}}} &
\multicolumn{1}{c}{\parbox[t]{2mm}{\multirow{5}{*}{\rotatebox[origin=c]{90}{PPTA}}}} &
\multicolumn{1}{c}{DM$_k$} &
\multicolumn{1}{c}{$\sigma_{w}$} &
\multicolumn{1}{c}{Span} &
\multicolumn{1}{c}{References } \\
\multicolumn{1}{c}{} &
\multicolumn{1}{c}{period} &
\multicolumn{1}{c}{(cm$^{-3}$~pc)} &
\multicolumn{1}{r}{(mJy)} &
\multicolumn{1}{c}{(kpc)} &
\multicolumn{1}{c}{} &
\multicolumn{1}{c}{} &
\multicolumn{1}{c}{} &
\multicolumn{1}{c}{} &
\multicolumn{1}{c}{($\mu$s)} &
\multicolumn{1}{c}{(year)} &
\multicolumn{1}{c}{}  \\
\multicolumn{1}{c}{} &
\multicolumn{1}{c}{(ms)} &
\multicolumn{1}{c}{ } &
\multicolumn{1}{r}{} &
\multicolumn{1}{c}{ } &
\multicolumn{1}{c}{} &
\multicolumn{1}{c}{} &
\multicolumn{1}{c}{} &
\multicolumn{1}{c}{} &
\multicolumn{1}{c}{} &
\multicolumn{1}{c}{} &
\multicolumn{1}{c}{}  \\\\\\
\hline
J1824$-$2452A & 3.054 & 119.89 & 2.0 & $3.7\pm0.7$ & & & X & Y & 0.57 & 5.6 & 46, 14, 21 \\
J1832$-$0836$^*$ & 2.719 & 28.18 & 1.1 & $0.8\pm0.2$ & & X & & Y & 1.86 & 0.6 & 47, 2 \\
J1843$-$1113 & 1.846  & 59.96 & 0.1 & $1.7\pm0.3$ & X & & & N & 0.71 & 10.0 & 48, 9 \\
J1853$+$1303 & 4.092 & 30.57 & 0.4 & $1.3\pm0.3$ & X & X & & Y & 1.31 & 8.4 & 43, 2, 42  \\
J1857$+$0943 & 5.362 & 13.30 & 5.0 & $1.1\pm0.1\dagger$ & X & X & X & Y & 1.16 & 28.4 & 49, 2, 21 \\
J1903$+$0327$^*$ & 2.150 & 297.52 & 1.3 & $6.1\pm1.2$ & & X & & Y & 2.11 & 4.0 & 50, 51 \\
J1909$-$3744 & 2.947 & 10.39 & 2.1 & $1.14\pm0.01\dagger$ & X & X & X & Y & 0.19 & 10.8 & 52, 14, 16 \\
J1910$+$1256 & 4.984 & 38.07 & 0.5 & $1.5\pm0.3$ & X & X & & Y & 1.42 & 9.5 & 43, 2, 42 \\
J1911$-$1114 & 3.626 & 31.02 & 0.5 & $1.1\pm0.2$ & X & & & N & 4.30 & 7.5 & 33, 9, 11 \\
J1911$+$1347 & 4.626 & 30.99 & 0.1 & $1.4\pm0.3$ & X & & & N & 1.09 & 8.8 & 43, 2, 44 \\
J1918$-$0642 & 7.646 & 26.55 & 0.6 & $1.3_{-0.1}^{+0.2}\dagger$ & X & X & & Y & 1.80 & 12.8 & 36, 2, 37 \\
J1923$+$2515$^*$ & 3.788 & 18.86 & 0.2 & $1.2\pm0.2$ & & X & & Y & 2.25 & 2.2 & 53, 2, 3 \\
J1939$+$2134 & 1.558 & 71.02 & 13.2 & $4.7_{-0.9}^{+1.4}\dagger$ & X & X & X & Y & 0.24 & 29.4 & 54, 9, 21 \\
J1944$+$0907$^*$ & 5.185 & 24.34 & 2.6 & $1.2\pm0.2$ & & X & & Y & 2.22 & 5.7 & 55, 2, 3 \\
J1949$+$3106$^*$ & 13.138 & 164.13 & 0.2 & $7.5\pm1.5$ & & X & & Y & 4.61 & 1.2 & 56 \\
J1955$+$2908 & 6.133 & 104.50 & 1.1 & $6.3\pm1.3$ & X & X & & Y & 3.20 & 8.1 & 57, 2, 11 \\
J2010$-$1323 & 5.223 & 22.16 & 1.6 & $1.9_{-0.5}^{+0.8}\ddag$ & X & X & & Y & 2.53 & 7.4 & 32, 2, 30 \\
J2017$+$0603$^*$ & 2.896 & 23.92 & 0.5 & $1.4\pm0.3$ & & X & & Y & 0.72 & 1.7 & 58, 2 \\
J2019$+$2425 & 3.934 & 17.20 & -- & $1.2\pm0.2$ & X & & & N & 9.64 & 9.1 & 59, 60 \\
J2033$+$1734 & 5.949 & 25.08 & -- & $1.7\pm0.3$ & X & & & N & 13.65 & 7.9 & 61, 2  \\
J2043$+$1711$^*$ & 2.380 & 20.71 & -- & $1.1\pm0.1\dagger$ & & X & & Y & 0.63 & 2.3 & 1, 2 \\
J2124$-$3358 & 4.931 & 4.60 & 3.6 & $0.39_{-0.04}^{+0.05}\dagger$ & X & & X & Y & 2.89 & 20.0 & 24, 14, 21 \\
J2129$-$5721 & 3.726 & 31.85 & 1.1 & $0.6_{-0.2}^{+0.6}\ddag$ & & & X & Y & 0.98 & 15.4 & 33, 14, 21, 31 \\
J2145$-$0750 & 16.052 & 9.00 & 8.9 & $0.62\pm0.02\ddag$ & X & X & X & Y & 1.73 & 21.2 & 5, 9, 14, 16, 30 \\
J2214$+$3000$^*$ & 3.119 & 22.55 & 0.5 & $0.9\pm0.2\dagger$ & & X & & Y & 1.67 & 2.1 & 62, 2, 3 \\
J2229$+$2643 & 2.978 & 22.72 & 0.9 & $1.8\pm0.4$ & X & & & N & 4.28 & 8.2 & 63, 2, 11 \\
J2302$+$4442$^*$ & 5.192 & 13.73 & 1.2 & $0.9\pm0.2$ & & X & & Y & 5.82 & 1.7 & 58, 27 \\
J2317$+$1439 & 3.445 & 21.90 & 4 & $0.7_{-0.3}^{+0.7}\ddag$ & X & X & & Y & 0.87 & 17.3 & 64, 2, 11, 30 \\
J2322$+$2057 & 4.808 & 13.36 & -- & $1.0\pm0.2$ & X & & & N & 6.74 & 7.9 & 59, 9  \\
\hline
\end{tabular}
\end{center}
\end{table*}

\section{Creating the IPTA dr2 data set}
\label{procedure}

We combine ToA measurements from individual PTA data releases into a single data set, and then perform the timing analysis for each pulsar in that data set.  In this work, we always use ToAs as reported in the individual PTA data sets, i.e., we have not re-processed raw observational data.  When possible, we include metadata (such as observation time and bandwidth) either as reported in the individual PTA data set or extracted from the original raw data files. In some cases \citep[e.g., early WSRT observations and the data reported in][]{ktr94}, it was not possible to recover a full set of metadata.

The data combination procedure is detailed in \citet{vlh+16}, and we summarise it here. We use the pulsar timing software package \textsc{tempo2}\footnote{https://bitbucket.org/psrsoft/tempo2} \citep{ehm06,hem06} to fit the timing model to the observed ToAs and obtain timing residuals (i.e. the difference between the observed and predicted ToAs) of the pulsar. We combine the different data sets of a given pulsar by fitting for time offsets (or `JUMPs') in the timing model to account for any systematic delays between them \citep[see][]{vlh+16}. We define the highest weighted data set (i.e. the sum of 1/$\sigma^2$, where $\sigma$ is the ToA uncertainty) as the reference data set (i.e. JUMP is equal to zero) in the timing model of the pulsar and then include separate JUMPs for each of other data sets to constrain their time offsets with respect to the reference. We note that, as mentioned in \S~\ref{data}, the PPTA data set contains several measured backend-dependent time offsets, which we include as fixed JUMPs in the timing model.

A pulsar timing model generally consists of astrometric parameters (right ascension RA, declination DEC, proper motion in RA and DEC, timing parallax $\pi$), rotational frequency information (spin frequency $f$ and its time derivatives), and dispersion measure information (DM; this accounts for the frequency-dependent time delay of the pulses due to electrons in the interstellar medium along the line-of-sight). With consistent adequate bandwidth and/or multiple observing frequencies, the time dependence of DM can also be included in the model. If the pulsar is in a binary system, the Keplerian parameters (orbital period $P_b$, projected semi-major axis $x$ of the pulsar orbit, longitude of periastron $\omega_0$, epoch of periastron passage $T_0$, and eccentricity $e$ of the orbit) are included to describe its binary motion. Some pulsars also require theory-independent Post-Keplerian parameters (orbital period derivative $\dot{P}_b$, periastron advance $\dot{\omega}_0$, Shapiro delay parameters `range $r$' and `shape $s$', apparent derivative of the projected semi-major axis $\dot{x}$) to account for any deviation of the orbit from Keplerian motion \citep[see][]{dd85,dd86,dt92}.
 A detailed description of all these parameters is given in \citet{lk05}. We use the \textsc{tempo2} binary model T2 in general in timing models of binary pulsars. For low-eccentricity pulsars, we use the binary model ELL1 \citep{wex99,lcw+01}, in which the first and second Laplace-Lagrange parameters ($\epsilon_1=e\sin\omega_0$ and $\epsilon_2=e\cos\omega_0$) are fitted. For low-eccentricity and medium- to high-inclination binary pulsars, we use the binary model DDH \citep{fw10} in the timing model in which the amplitude of the third harmonic of the orbital period (H3) and the ratio of amplitudes of successive harmonics (STIG) are fitted.
 
 In the fitting process, the measured topocentric ToAs are converted to the solar-system Barycentric Coordinate Time (TCB) through the solar-system ephemeris DE436\footnote{This solar-system ephemeris is based on \citet{fwb+14}.} using the Terrestrial Time standard BIPM2015\footnote{This time standard has been obtained according to principles given in \citet{gui88} and \citet{pet03b}.}. 
 The Barycentric Dynamical Time (TDB) is commonly used in astronomy and thus, we also convert our timing results to TDB units and include the solutions in the data release separately. To develop the timing model for each pulsar, we started by fitting the timing model parameters from one of the individual PTA data releases. We then added any additional parameters needed to accommodate the other individual PTA data releases, and we tested for further parameters that might be needed in the combined data set (as described below).

If the pulsar is observed by NANOGrav, we then include `frequency-dependent' (FD) parameters in the timing model because of the availability of the frequency-dependent sub-band ToAs to minimise the effect of frequency-dependent pulse profile evolution \citep{abb+15}. The number of required FD parameters for a given pulsar is obtained from \citet{abb+15}.

To model the white noise $\sigma$ (uncorrelated in time) of the pulsar data, we include the standard noise parameters EFAC (${\rm E_f}$) and EQUAD (${\rm E_q}$) for each data set in the timing model \citep[see][for details]{vlh+16}. 
The EFAC is a scale parameter on the ToA uncertainty and the EQUAD is an added variance that mainly accounts for the error caused by pulse phase-jitter \citep{ovh+11,ovd+13} and other systematic effects. The \textsc{tempo2} version \citep{ehm06} defines the relationship of these two parameters to a ToA uncertainty ${\rm \sigma_t}$ as

\begin{equation}
\label{t2}
\sigma = {\rm E_f} \sqrt{{\rm E_q^2} + {\rm \sigma_t^2}},
\end{equation}

\noindent
while the \textsc{temponest} version \citep{lah+14} defines the relationship in a reverse order as

\begin{equation}
\label{tn}
\sigma = \sqrt{{\rm E_q^2} + {\rm E_f^2 \sigma_t^2}}.
\end{equation}

\noindent
\textsc{temponest}\footnote{https://github.com/LindleyLentati/TempoNest} is a pulsar noise analysis plugin in \textsc{tempo2} that is based on Bayesian analysis \citep{lah+14}. We also include the factor ECORR to correct for the pulse phase jitter that causes correlation between simultaneous ToAs obtained at different observing frequencies. In the data combination, we include separate EFACs and EQUADs for all telescope/backend-dependent PTA data sets, and separate ECORRs for telescope/backend-dependent NANOGrav data sets because of their available simultaneous frequency-dependent ToAs (i.e. sub-band ToAs).

In addition to white noise, we model the time-correlated red noise processes by including the stochastic DM variation and the spin noise processes with power-law models in the timing solution. We use the \textsc{temponest} plugin to determine the white and red noise parameters by fitting simultaneously while marginalising over the timing model parameters \citep[see][for more details of the software]{lah+14}. We note that, as shown in \citet{lsc+16}, some pulsars in the IPTA dr1 needed additional red noise processes such as ``system noise'' and ``band noise'' to accurately model the noise in their timing data. 
The system noise models possible instrumental effects and calibration errors that might appear in a single observing system or telescope. The band noise models signals that exist in a given frequency band. These signals may have originated in the interstellar medium due to processes that are incoherent between different bands, or that do not scale in amplitude with the inverse square of the observing frequency, or due to radio frequency interference that present in the same band independent of the observing site.
Therefore, our basic DM and spin noise processes may not provide the optimal model. A detailed noise analysis will be carried out separately in a future study.

The individual PTA data releases used different methods of modelling DM variations: the EPTA data release used the first two time derivatives of the DM  and the stochastic DM variation with a power-law \citep{dcl+16,cll+16}; the PPTA data release used only the first two time derivatives of the DM \citep{rhc+16}; and the NANOGrav data release measured the change in DM, relative to the fiducial DM value in the timing model, at nearly every observing epoch using the {\sc tempo}\footnote{\url{http://tempo.sourceforge.net/}} `DMX' parameter \citep{abb+15}. In this IPTA data combination, we use two different methods to model the DM variation, including the model DMMODEL given in \citet{kcs+13} as described in \S~\ref{verA}, and time derivatives of the DM with a power-law stochastic DM variation as described in \S~\ref{verB}.

Finally, with the fully combined data set and timing model for each pulsar, we use an $F$-test with the residual sum of squares of each model, as described in \citet{abb+15}, to search for parameters that have become significant as a result of combining the data. This process is used in all NANOGrav data releases to ensure the model is as complete as possible. With the addition of a new parameter, an 
$F$-test significance value of $\leq 0.0027$ (i.e., $3\sigma$ significance) implies that the additional parameter has significantly improved the model's description of the data. For long data sets, the most likely parameters to become significant are post-Keplerian parameters; additionally, the use of wide-bandwidth or multi-band data may require higher-order ``FD'' parameters to model frequency-dependent pulse shape evolution, as described in \citet{abb+15}. After having combined the timing models as described earlier in this section, with this $F$-test analysis we do not find any additional parameters that are required in IPTA dr2 pulsar timing models beyond those used in the individual PTA data sets.

\section{Results}
\label{results}

We produce two data combination versions (VersionA and VersionB) in the IPTA dr2 and the data set is available at {\url{http://www.ipta4gw.org}}. We also note that the data set includes separate timing solutions for pulsars produced with TCB and TDB units. The two versions are different in terms of modelling the DM variation and handling the noise properties of pulsars, and they are described below in detail.

\subsection{IPTA dr2 -- VersionA}
\label{verA}

In this version, we determine the DM variation using the model given in \citet{kcs+13} and implemented in \textsc{tempo2} as DMMODEL. This model estimates the DM offsets from the global value as a function of time for a given time grid. We use a 60~day MJD grid in general for all pulsars in the combination, but a 30~day grid is used for several sources to better constrain DM variations. For pulsars with a lack of multi-frequency observations (or a shorter time span of multi-frequency coverage), the DMMODEL does not provide reliable results and thus, we use the basic time derivatives of the DM (i.e. $\dot{\rm DM}$ and $\ddot{\rm DM}$) in the timing model (see the ninth column in Table~\ref{psrs_par}). We include only white noise parameters EFACs and EQUADs in the timing model of pulsars in this version. Note that we do not constrain them using this IPTA data combination, rather we use the values constrained in previous data releases. The EFACs and EQUADs for the EPTA data are taken from the EPTA dr1 \citep{dcl+16}, those for the PPTA data are taken from the IPTA dr1 \citep{vlh+16}. These EPTA and PPTA white noise parameters were constrained using \textsc{temponest} according to the ToA uncertainty scaling given in Equation~\ref{tn}. For NANOGrav data, we use the \textsc{tempo2} version of EFACs and EQUADs (see Equation~\ref{t2}), which are taken from the NANOGrav data release \citep{abb+15}. Finally we update timing models of all 65 pulsars by running \textsc{tempo2} using the combined IPTA data set.

\begin{figure*}
\includegraphics[width=17cm]{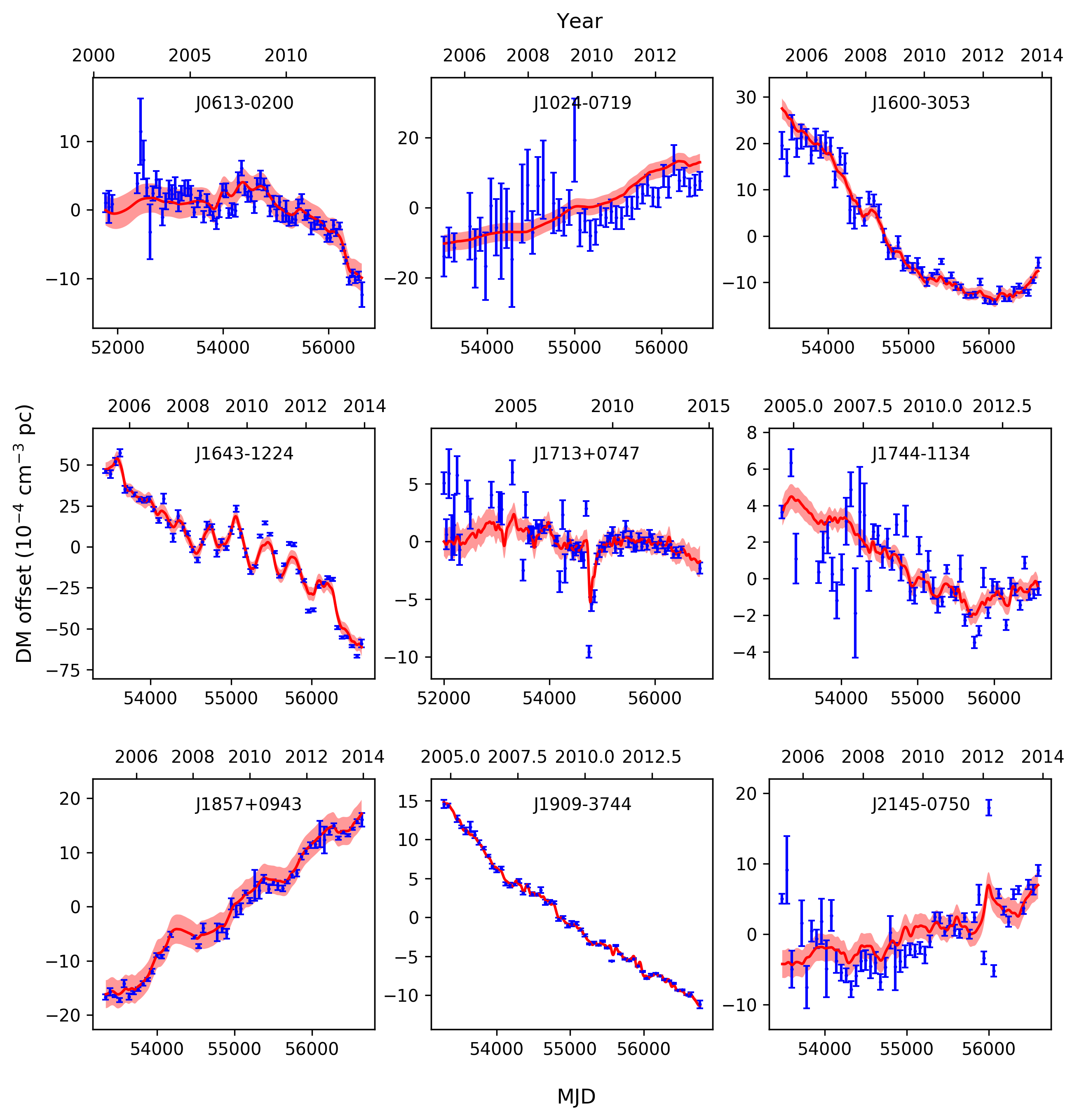}
\caption{
The time-dependent DM variations for pulsars that are observed by all three PTAs obtained using the DMMODEL ({\it blue}) as described in VersionA (see \S~\ref{verA}). Note that the mean DM is subtracted and only the variation is plotted. For comparison, the DM variations obtained using the power-law model described in VersionB (see \S~\ref{verB}) are over-plotted ({\it red}), and their uncertainties are estimated using a Gaussian process regression method. 
The overall features in DM variations obtained from the two versions are consistent with each other within their measured uncertainties. Note that we omitted PSR J1939$+$2134 in this figure because of its complicated timing noise behaviour. 
}
\label{dm_curves}
\end{figure*}

Figure~\ref{dm_curves} shows the time-dependent DM variation obtained from DMMODEL for pulsars that are observed by all three PTAs. Note that we did not include PSR J1939$+$2134 in Figure~\ref{dm_curves} because of its complicated DM variation and timing noise \citep[e.g. ][]{ktr94,mhb+13,abb+15,dcl+16,cll+16,lsc+16}. These results are consistent with the DM variations of pulsars presented in \citet{kcs+13} using the PPTA data, and also with the results obtained using the DMX method that are presented in \citet{abb+15} using the NANOGrav data.

\subsection{IPTA dr2 -- VersionB}
\label{verB}

The main difference of this version compared to VersionA is in the modelling of the white and red noise processes and DM variations of the pulsars.
We re-estimate all the noise parameters of pulsars based on this IPTA data combination, rather than using previously constrained values given in other data releases. We include new EFACs and EQUADs for all PTA data sets and separate ECORRs for NANOGrav data sets if available in the pulsar timing model. We include the first two time derivatives of the DM and then model the time-correlated stochastic DM and the red spin noise processes using separate power-law models in the timing model. Using \textsc{temponest}, we then constrain these noise parameters simultaneously while marginalising over the timing model parameters. For comparison with VersionA, we over-plot the DM variations for pulsars that are observed by all three PTAs in Figure~\ref{dm_curves}. This shows that the overall time-dependent DM variations modeled by these two methods are largely consistent with each other within their uncertainties. 

\begin{figure*}
\includegraphics[width=17.5cm]{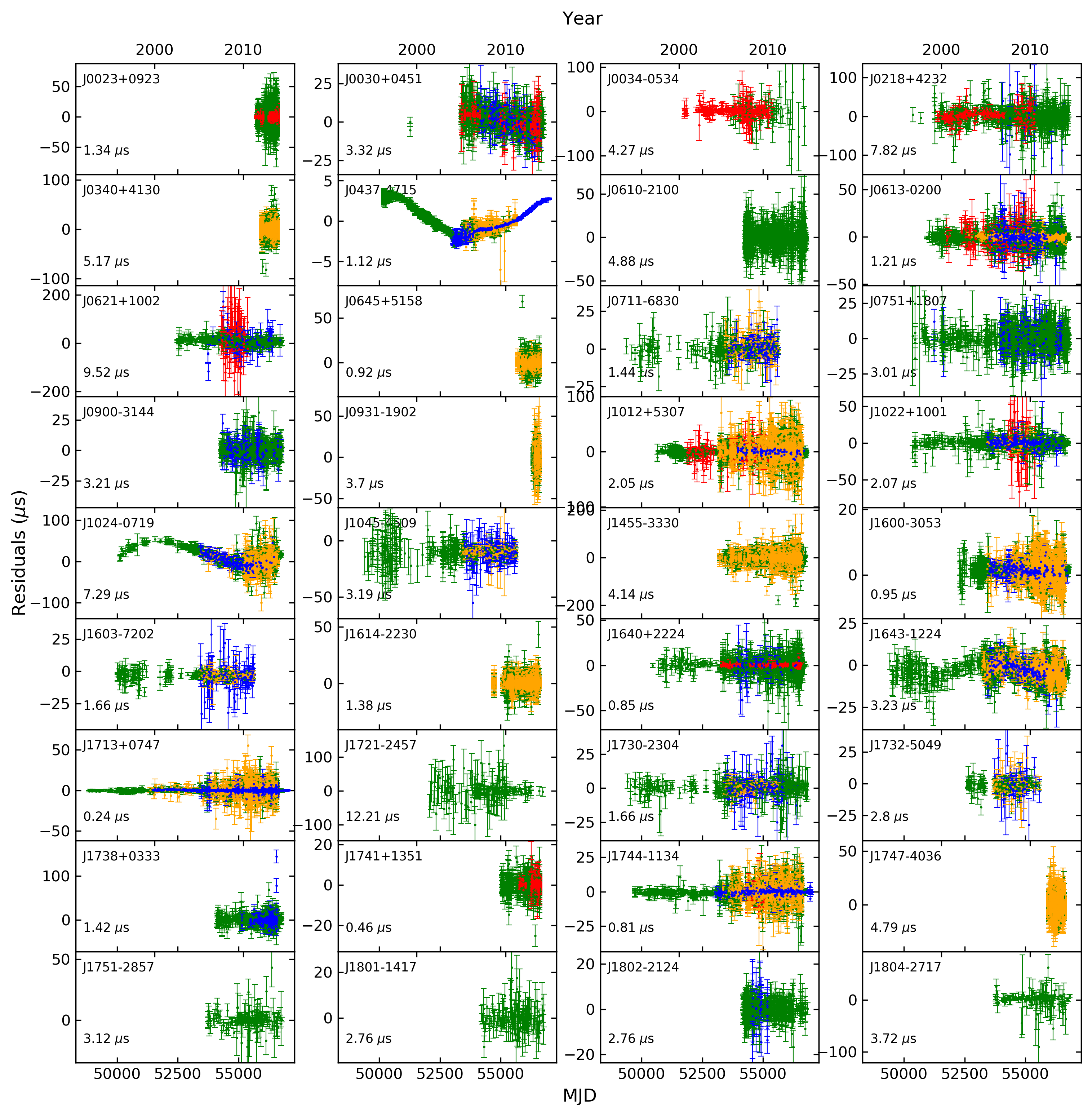}
\caption{
The timing residuals (i.e. the difference between the observed and timing model predicted ToAs) of the first 36 pulsars obtained using the data combination VersionB (see \S~\ref{verB}). The maximum-likelihood waveform of the power-law stochastic DM variation model is subtracted from the residuals, but the red spin noise model has not been subtracted. The pulsar name is given in the top and the weighted root-mean-square of the timing residuals is given in the bottom of each panel. The colour-code represents different observing frequencies as given in Figure~\ref{psrs_freqs}: $<$0.5~GHz (red), 0.5--1.0~GHz (orange), 1--2~GHz (green), $>$2~GHz (blue). 
}
\label{residual_1}
\end{figure*}

\begin{figure*}
\includegraphics[width=17.5cm]{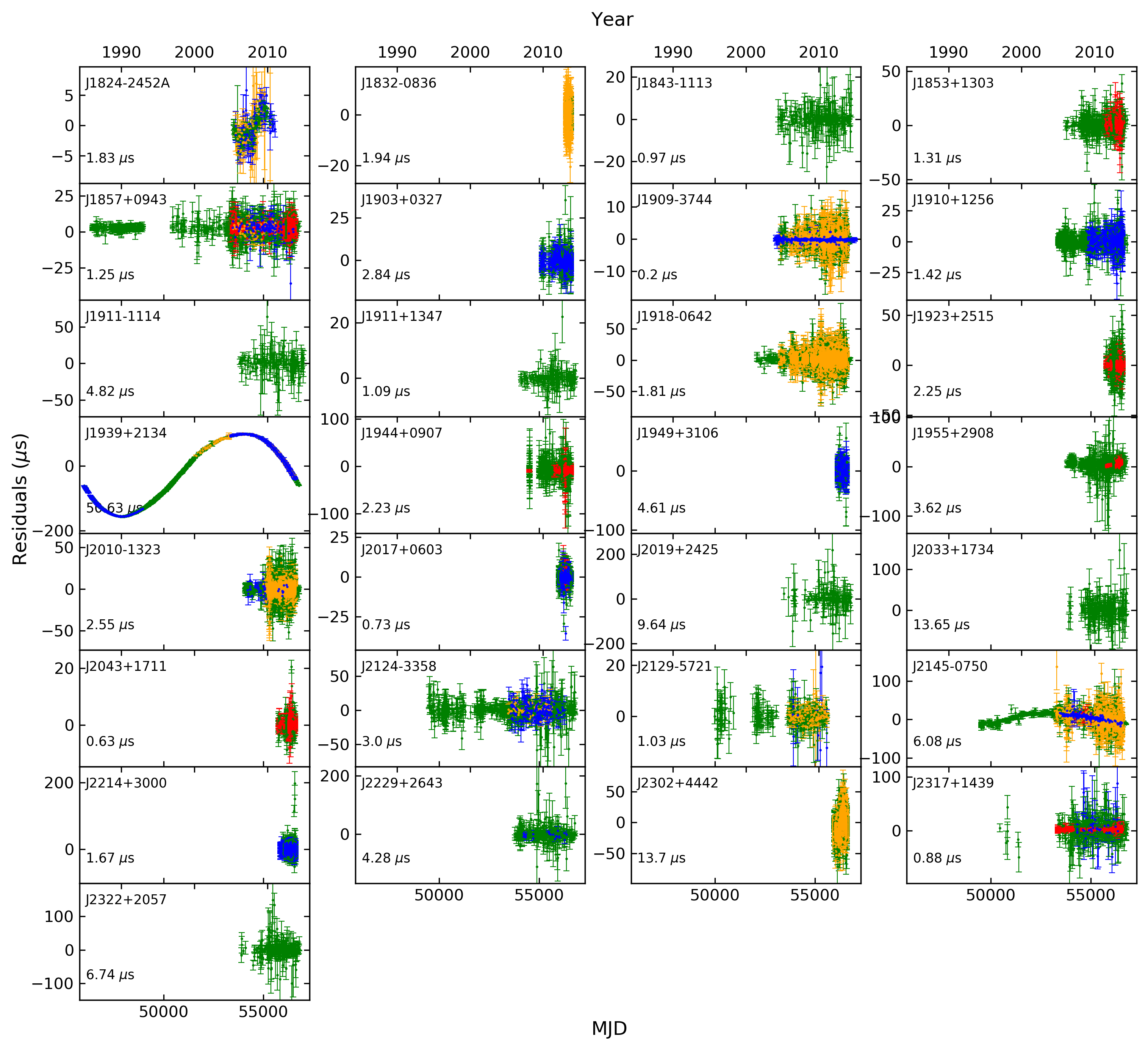}
\caption{
Same as Figure~\ref{residual_1}, but for the last 29 pulsars.
}
\label{residual_2}
\end{figure*}

We present the timing residuals of pulsars in Figure~\ref{residual_1} and Figure~\ref{residual_2}. We have subtracted the power-law waveform of the DM stochastic noise in these residuals, but not the waveforms of red spin noise processes. Some pulsars exhibit complicated noise processes and need a more sophisticated noise analysis including various additional noise terms such as systematic noise and band noise as discussed in \citet{lsc+16}. This will be done separately combining with GW search analyses using this new data combination in the future. We present the best timing models for all our pulsars in Appendix~\ref{app}.

\section{Discussion}
\label{dis}

In this paper, we presented the creation of the IPTA second data release (IPTA dr2) that includes the EPTA, NANOGrav, and PPTA data releases presented in \citet{dcl+16}, \citet{abb+15}, and \citet{rhc+16}, respectively. This new IPTA data release consists of regularly observed high-precision timing data of 65 MSPs, which includes 16 additional MSPs compared to the previous IPTA dr1. We produced two versions in the data release (i.e. VersionA and VersionB) depending on different methods of handling the DM and the noise processes of pulsars as described in Section~\ref{results}. 
We directly compared the timing ephemerides of pulsars obtained from the two versions in this new IPTA dr2 and the previous IPTA dr1. We found that all the timing parameters are greatly consistent with each other and their uncertainties resulted in IPTA dr2 are generally improved compared to the IPTA dr1, mostly due to the addition of more data in the combination. 
We also compared the DM variations of pulsars obtained using VersionA and VersionB (see Figure~\ref{dm_curves}). This comparison shows that the overall features in the variations are consistent with each other within their uncertainties, including the uncertainty of the mean DM measurement in the timing model. We note that the DMMODEL provides much noisier variation compared to the power-law model. This is because the DMMODEL follows a piece-wise method using a given time-grid, which depends on observation sampling and the availability of multi-frequency data \citep[see][]{kcs+13}. In contrast, the power-law model fits for the power spectrum of the timing data and the waveform of the DM variation can be generated for any given time series \citep[see][]{lah+14,lbj+14}.

We only constrained the basic noise properties of pulsars in this data combination. However, using IPTA dr1, \citet{lsc+16} showed that some pulsars need additional noise terms such as system and band noise to model their overall noise properties accurately because of the involvement of several observing systems (i.e. backends/telescopes) in the observations and also wider frequency coverages. 
By simply comparing the weighted root mean square (rms) of timing residuals after subtracting the maximum-likelihood time-correlated noise signals \citep[i.e., by comparing $\sigma_w$ in Table~\ref{psrs_par} herein and Table~1 in][]{lsc+16}, we found that approximately 60 per cent of pulsars have improved their timing precision based on this new IPTA dr2 compared to the previous IPTA dr1.
The rest of the pulsars have slightly poorer timing precision compared to the previous data release, probably because these pulsars require additional noise terms to optimise their noise analysis as described in \citet{lsc+16} which we have ignored in the present analysis. Thus, a detailed noise modeling based on the IPTA dr2 will be required and conducted in subsequent analysis. This will be published separately in the future.

Adding new data is essential to improve the timing precision and the sensitivity of the IPTA to GWs. We also need to consider and understand these new data and their noise behaviour to be able to achieve optimal results. This requires additional noise investigation and more computationally expensive methods to model their noise behaviours, which can be very time consuming.
For instance, the data sets of PSRs J1713$+$0747 and J1939$+$2134 in the IPTA dr2 are long and dense due to the involvement of all IPTA telescopes in the observations with several backends providing broad frequency coverage. Based on our basic noise analysis, J1713$+$0747 and J1939$+$2134 required 90 and 74 noise parameters, respectively, in the timing model requiring weeks of computer time to conduct their noise analyses. We will have more pulsar data available for the IPTA in the future and their noise analyses will become even more complicated. Therefore, we need to investigate methods to improve the efficiencies of current pulsar noise analysis software and also novel techniques to constrain noise in more efficient ways. While we have assumed here that all stochastic processes have power-law spectra, in the future it may necessary to consider more complex models to be able to understand the pulsar noise behaviours more accurately. This will become important especially when high-resolution data are available for the IPTA from modern telescopes such as MeerKAT and Square Kilometre Array, and also with the instrument upgrades of current telescopes in the IPTA.

\section{Future IPTA studies}
\label{future}

The primary goal of the IPTA is to detect and characterize low-frequency GWs using high-precision pulsar timing \citep{vlh+16}. The IPTA dr2 is the most complete MSP data set produced up-to-date for GW search experiments. There are a suite of papers which are currently exploring the broader impacts of IPTA dr2. In terms of GW search analyses, we are preparing improved GW background constraints that revise upper limits from \cite{srl+15} and \cite{ltm+15} by accounting for solar-system effects using \textsc{BayesEphem} \citep{abbGWB18}, 
and will apply more flexible DM variation models to \cite{abb+15}. We are exploring the detection response of the IPTA to the GW background, and how this compares to that of the constituent regional PTA data sets. We are also carrying out a search for GW memory \citep{bt87, vHL10, abb15,whc+15, mcc17}, in addition to exploring new and novel ways of analyzing IPTA data. These include (but are not limited to) the identification and removal of legacy ToAs which do not contribute to our GW background sensitivity, as well as the preparation of smaller IPTA data sets that require minimal combination efforts from the constitutent PTAs, thereby enabling fast diagnostics. Potential avenues of future GW study with this data include searching for individually-resolvable supermassive black hole binary systems \citep{zhw+14,bps+16,aab+18}, and placing constraints on beyond-General-Relativity GW polarization states \citep{ljp08,ljp+10,cs12,lee13,grt15,cot+18,ocv+19}.

We expect IPTA dr2 to also impact many areas that are synergistic to nanohertz GW searches, including (but not limited to): probing ultralight scalar-field dark matter (the so-called ``fuzzy'' dark matter) in the particle mass range $10^{-24}-10^{-22}$ eV \citep[][and references therein]{pzl+18}; improving the characterization of radio-frequency--dependent delay processes induced by the ionized interstellar medium \citep{kcs+13,lcc+17,jml+17} and solar wind \citep{mca+19,nhw+17,tvs+19}; studying the solar system and giving independent constraints on ephemeris parameters with pulsar-timing data \citep{abbGWB18,glc18,cgl+18}; and synthesizing a pulsar-based time standard \citep{hcm+12}. Several of these goals (both GW and synergistic) may be aided by improved pulsar distance precision and discovery techniques \citep[e.g.][]{dgb+19,mab+19,jkc+18}.

\section*{Acknowledgements}

The National Radio Astronomy Observatory is a facility of the National Science Foundation operated under cooperative agreement by Associated Universities, Inc.
The NANOGrav Physics Frontiers Center is supported by NSF award number 1430284. The Green Bank Observatory is a facility of the National Science Foundation operated under cooperative agreement by Associated Universities, Inc.
The Arecibo Observatory is operated by the University of Central Florida, Ana G. Mendez-Universidad Metropolitana, and Yang Enterprises under a cooperative agreement with the National Science Foundation (NSF; AST-1744119). 
The Parkes telescope is part of the Australia Telescope which is funded by the Commonwealth Government for operation as a National Facility managed by CSIRO. The Westerbork Synthesis Radio Telescope is operated by the Netherlands Institute for Radio Astronomy (ASTRON) with support from The Netherlands Foundation for Scientific Research NWO. The 100-m Effelsberg Radio Telescope is operated by the Max-Planck-Institut f\"{u}r Radioastronomie at Effelsberg. Some of the work reported in this paper was supported by the ERC Advanced Grant `LEAP', Grant Agreement Number 227947 (PI Kramer). Pulsar research at the Jodrell Bank Centre for Astrophysics is supported by a consolidated grant from STFC. The Nancay radio telescope is operated by the Paris Observatory, associated with the Centre National de la Recherche Scientifique (CNRS) and acknowledges financial support from the `Programme National de Cosmologie et Galaxies (PNCG)' and `Gravitation, R\'{e}f\'{e}rences, Astronomie, M\'{e}trologie (GRAM)' programmes of CNRS/INSU, France. The Flatiron Institute is supported by the Simons Foundation. This work was supported by the NANOGrav Physics Frontiers Center (NSF award 1430284).
Parts of this research were conducted by the Australian Research Council Centre of Excellence for Gravitational Wave Discovery (OzGrav), through project number CE170100004.
SO and RS acknowledge Australian Research Council grant FL150100148.
SMR is a CIFAR Fellow.
GD and KL acknowledges financial support by the European Research Council for the ERC Synergy Grant BlackHoleCam under contract no. 610058.
KJL is supported by XDB23010200, NSFC U15311243, 2017YFA0402600 and funding from TianShanChuangXinTuanDui and Max-Planck Partner Group.
MAM and SBS are supported by NSF award number 1458952.
JBW is supported by the Youth Innovation Promotion Association of Chinese Academy of Sciences.
WWZ is supported by Chinese Academy of Science Pioneer Hundred Talents Program, the Strategic Priority Research Program of the Chinese Academy of Sciences Grant No. XDB23000000, and by the National Natural Science Foundation of China under grant No. 11743002, 11873067.
Part of this research was performed at the Jet Propulsion Laboratory, under contract with the National Aeronautics and Space Administration. JS and MV acknowledges support from the JPL R\&TD program.
Basic pulsar research at NRL is funded by the Chief of Naval Research.
Pulsar research at UBC is supported by an NSERC Discovery Grant and by the Canadian Institute for Advanced Research.

\bibliography{psrrefs,modrefs,journals}

\begin{thebibliography}{}
\makeatletter
\relax
\def\mn@urlcharsother{\let\do\@makeother \do\$\do\&\do\#\do\^\do\_\do\%\do\~}
\def\mn@doi{\begingroup\mn@urlcharsother \@ifnextchar [ {\mn@doi@}
  {\mn@doi@[]}}
\def\mn@doi@[#1]#2{\def\@tempa{#1}\ifx\@tempa\@empty \href
  {http://dx.doi.org/#2} {doi:#2}\else \href {http://dx.doi.org/#2} {#1}\fi
  \endgroup}
\def\mn@eprint#1#2{\mn@eprint@#1:#2::\@nil}
\def\mn@eprint@arXiv#1{\href {http://arxiv.org/abs/#1} {{\tt arXiv:#1}}}
\def\mn@eprint@dblp#1{\href {http://dblp.uni-trier.de/rec/bibtex/#1.xml}
  {dblp:#1}}
\def\mn@eprint@#1:#2:#3:#4\@nil{\def\@tempa {#1}\def\@tempb {#2}\def\@tempc
  {#3}\ifx \@tempc \@empty \let \@tempc \@tempb \let \@tempb \@tempa \fi \ifx
  \@tempb \@empty \def\@tempb {arXiv}\fi \@ifundefined
  {mn@eprint@\@tempb}{\@tempb:\@tempc}{\expandafter \expandafter \csname
  mn@eprint@\@tempb\endcsname \expandafter{\@tempc}}}

\bibitem[\protect\citeauthoryear{{Abdo} et~al.,}{{Abdo} et~al.}{2010}]{aaa+10e}
{Abdo} A.~A.,  et~al., 2010, \mn@doi [\apj] {10.1088/0004-637X/712/2/957},
  \href {http://adsabs.harvard.edu/abs/2010ApJ...712..957A} {712, 957}

\bibitem[\protect\citeauthoryear{{Aggarwal} et~al.,}{{Aggarwal}
  et~al.}{2018}]{aab+18}
{Aggarwal} K.,  et~al., 2018, arXiv e-prints, \href
  {https://ui.adsabs.harvard.edu/abs/2018arXiv181211585A} {p. arXiv:1812.11585}

\bibitem[\protect\citeauthoryear{Alpar, Cheng, Ruderman  \& Shaham}{Alpar
  et~al.}{1982}]{acrs82}
Alpar M.~A.,  Cheng A.~F.,  Ruderman M.~A.,   Shaham J.,  1982, \nat, 300, 728

\bibitem[\protect\citeauthoryear{{Arzoumanian} et~al.,}{{Arzoumanian}
  et~al.}{2014}]{abb+14}
{Arzoumanian} Z.,  et~al., 2014, \mn@doi [\apj] {10.1088/0004-637X/794/2/141},
  \href {http://adsabs.harvard.edu/abs/2014ApJ...794..141A} {794, 141}

\bibitem[\protect\citeauthoryear{{Arzoumanian} et~al.,}{{Arzoumanian}
  et~al.}{2015a}]{abb15}
{Arzoumanian} Z.,  et~al., 2015a, \mn@doi [\apj] {10.1088/0004-637X/810/2/150},
  \href {http://adsabs.harvard.edu/abs/2015ApJ...810..150A} {810, 150}

\bibitem[\protect\citeauthoryear{{Arzoumanian} et~al.,}{{Arzoumanian}
  et~al.}{2015b}]{abb+15}
{Arzoumanian} Z.,  et~al., 2015b, \mn@doi [\apj] {10.1088/0004-637X/813/1/65},
  \href {http://adsabs.harvard.edu/abs/2015ApJ...813...65T} {813, 65}

\bibitem[\protect\citeauthoryear{{Arzoumanian} et~al.,}{{Arzoumanian}
  et~al.}{2016}]{abb+16}
{Arzoumanian} Z.,  et~al., 2016, \mn@doi [\apj] {10.3847/0004-637X/821/1/13},
  \href {http://adsabs.harvard.edu/abs/2016ApJ...821...13A} {821, 13}

\bibitem[\protect\citeauthoryear{{Arzoumanian} et~al.,}{{Arzoumanian}
  et~al.}{2018a}]{abb+18}
{Arzoumanian} Z.,  et~al., 2018a, \mn@doi [\apjs] {10.3847/1538-4365/aab5b0},
  \href {http://adsabs.harvard.edu/abs/2018ApJS..235...37A} {235, 37}

\bibitem[\protect\citeauthoryear{{Arzoumanian} et~al.,}{{Arzoumanian}
  et~al.}{2018b}]{abbGWB18}
{Arzoumanian} Z.,  et~al., 2018b, \mn@doi [\apj] {10.3847/1538-4357/aabd3b},
  \href {http://adsabs.harvard.edu/abs/2018ApJ...859...47A} {859, 47}

\bibitem[\protect\citeauthoryear{{Babak} et~al.,}{{Babak}
  et~al.}{2016}]{bps+16}
{Babak} S.,  et~al., 2016, \mn@doi [\mnras] {10.1093/mnras/stv2092}, \href
  {http://adsabs.harvard.edu/abs/2016MNRAS.455.1665B} {455, 1665}

\bibitem[\protect\citeauthoryear{Backer, Kulkarni, Heiles, Davis  \&
  Goss}{Backer et~al.}{1982}]{bkh+82}
Backer D.~C.,  Kulkarni S.~R.,  Heiles C.,  Davis M.~M.,   Goss W.~M.,  1982,
  New~Astron., 300, 615

\bibitem[\protect\citeauthoryear{Bailes et~al.,}{Bailes et~al.}{1994}]{bhl+94}
Bailes M.,  et~al., 1994, ApJ, 425, L41

\bibitem[\protect\citeauthoryear{Bailes et~al.,}{Bailes et~al.}{1997}]{bjb+97}
Bailes M.,  et~al., 1997, ApJ, 481, 386

\bibitem[\protect\citeauthoryear{Boriakoff, Buccheri  \& Fauci}{Boriakoff
  et~al.}{1983}]{bbf83}
Boriakoff V.,  Buccheri R.,   Fauci F.,  1983, New~Astron., 304, 417

\bibitem[\protect\citeauthoryear{{Braginskii} \& {Thorne}}{{Braginskii} \&
  {Thorne}}{1987}]{bt87}
{Braginskii} V.~B.,  {Thorne} K.~S.,  1987, \mn@doi [\nat] {10.1038/327123a0},
  \href {http://adsabs.harvard.edu/abs/1987Natur.327..123B} {327, 123}

\bibitem[\protect\citeauthoryear{{Burgay} et~al.,}{{Burgay}
  et~al.}{2006}]{bjd+06}
{Burgay} M.,  et~al., 2006, MNRAS, 368, 283

\bibitem[\protect\citeauthoryear{{Burgay} et~al.,}{{Burgay}
  et~al.}{2013}]{bbb+13}
{Burgay} M.,  et~al., 2013, \mn@doi [\mnras] {10.1093/mnras/stt721}, \href
  {http://adsabs.harvard.edu/abs/2013MNRAS.433..259B} {433, 259}

\bibitem[\protect\citeauthoryear{{Caballero} et~al.,}{{Caballero}
  et~al.}{2016}]{cll+16}
{Caballero} R.~N.,  et~al., 2016, \mn@doi [\mnras] {10.1093/mnras/stw179},
  \href {http://adsabs.harvard.edu/abs/2016MNRAS.457.4421C} {457, 4421}

\bibitem[\protect\citeauthoryear{{Caballero} et~al.,}{{Caballero}
  et~al.}{2018}]{cgl+18}
{Caballero} R.~N.,  et~al., 2018, \mn@doi [\mnras] {10.1093/mnras/sty2632},
  \href {https://ui.adsabs.harvard.edu/abs/2018MNRAS.481.5501C} {481, 5501}

\bibitem[\protect\citeauthoryear{Camilo}{Camilo}{1995}]{cam95a}
Camilo F.,  1995, PhD thesis, Princeton University

\bibitem[\protect\citeauthoryear{Camilo, Nice  \& Taylor}{Camilo
  et~al.}{1993}]{cnt93}
Camilo F.,  Nice D.~J.,   Taylor J.~H.,  1993, ApJ, 412, L37

\bibitem[\protect\citeauthoryear{Camilo, Nice, Shrauner  \& Taylor}{Camilo
  et~al.}{1996}]{cnst96}
Camilo F.,  Nice D.~J.,  Shrauner J.~A.,   Taylor J.~H.,  1996, ApJ, 469, 819

\bibitem[\protect\citeauthoryear{{Camilo} et~al.,}{{Camilo}
  et~al.}{2015}]{ckr+15}
{Camilo} F.,  et~al., 2015, \mn@doi [\apj] {10.1088/0004-637X/810/2/85}, \href
  {http://adsabs.harvard.edu/abs/2015ApJ...810...85C} {810, 85}

\bibitem[\protect\citeauthoryear{{Chamberlin} \& {Siemens}}{{Chamberlin} \&
  {Siemens}}{2012}]{cs12}
{Chamberlin} S.~J.,  {Siemens} X.,  2012, \mn@doi [\prd]
  {10.1103/PhysRevD.85.082001}, \href
  {http://adsabs.harvard.edu/abs/2012PhRvD..85h2001C} {85, 082001}

\bibitem[\protect\citeauthoryear{{Champion} et~al.,}{{Champion}
  et~al.}{2005}]{clm+05}
{Champion} D.~J.,  et~al., 2005, MNRAS, 363, 929

\bibitem[\protect\citeauthoryear{{Champion} et~al.,}{{Champion}
  et~al.}{2008}]{crl+08}
{Champion} D.~J.,  et~al., 2008, \mn@doi [Science] {10.1126/science.1157580},
  \href {http://adsabs.harvard.edu/abs/2008Sci...320.1309C} {320, 1309}

\bibitem[\protect\citeauthoryear{{Cognard} et~al.,}{{Cognard}
  et~al.}{2011}]{cgj+11}
{Cognard} I.,  et~al., 2011, \mn@doi [\apj] {10.1088/0004-637X/732/1/47}, \href
  {http://adsabs.harvard.edu/abs/2011ApJ...732...47C} {732, 47}

\bibitem[\protect\citeauthoryear{{Corbin} \& {Cornish}}{{Corbin} \&
  {Cornish}}{2010}]{cc10}
{Corbin} V.,  {Cornish} N.~J.,  2010, arXiv e-prints, \href
  {https://ui.adsabs.harvard.edu/abs/2010arXiv1008.1782C} {p. arXiv:1008.1782}

\bibitem[\protect\citeauthoryear{{Cornish}, {O'Beirne}, {Taylor}  \&
  {Yunes}}{{Cornish} et~al.}{2018}]{cot+18}
{Cornish} N.~J.,  {O'Beirne} L.,  {Taylor} S.~R.,   {Yunes} N.,  2018, \mn@doi
  [Physical Review Letters] {10.1103/PhysRevLett.120.181101}, \href
  {http://adsabs.harvard.edu/abs/2018PhRvL.120r1101C} {120, 181101}

\bibitem[\protect\citeauthoryear{{Crawford}, {Roberts}, {Hessels}, {Ransom},
  {Livingstone}, {Tam}  \& {Kaspi}}{{Crawford} et~al.}{2006}]{crh+06a}
{Crawford} F.,  {Roberts} M.~S.~E.,  {Hessels} J.~W.~T.,  {Ransom} S.~M.,
  {Livingstone} M.,  {Tam} C.~R.,   {Kaspi} V.~M.,  2006, \mn@doi [\apj]
  {10.1086/508403}, \href {http://adsabs.harvard.edu/abs/2006ApJ...652.1499C}
  {652, 1499}

\bibitem[\protect\citeauthoryear{{Damour} \& {Deruelle}}{{Damour} \&
  {Deruelle}}{1985}]{dd85}
{Damour} T.,  {Deruelle} N.,  1985, Ann. Inst. Henri Poincar{\'e} Phys.
  Th{\'e}or, \href {https://ui.adsabs.harvard.edu/abs/1985AIHS...43..107D} {43,
  107}

\bibitem[\protect\citeauthoryear{{Damour} \& {Deruelle}}{{Damour} \&
  {Deruelle}}{1986}]{dd86}
{Damour} T.,  {Deruelle} N.,  1986, Ann. Inst. Henri Poincar{\'e} Phys.
  Th{\'e}or, \href {https://ui.adsabs.harvard.edu/abs/1986AIHS...44..263D} {44,
  263}

\bibitem[\protect\citeauthoryear{{Damour} \& {Taylor}}{{Damour} \&
  {Taylor}}{1992}]{dt92}
{Damour} T.,  {Taylor} J.~H.,  1992, \mn@doi [\prd] {10.1103/PhysRevD.45.1840},
  \href {https://ui.adsabs.harvard.edu/abs/1992PhRvD..45.1840D} {45, 1840}

\bibitem[\protect\citeauthoryear{{Deller}, {Verbiest}, {Tingay}  \&
  {Bailes}}{{Deller} et~al.}{2008}]{dvtb08}
{Deller} A.~T.,  {Verbiest} J.~P.~W.,  {Tingay} S.~J.,   {Bailes} M.,  2008,
  \mn@doi [\apjl] {10.1086/592401}, \href
  {https://ui.adsabs.harvard.edu/abs/2008ApJ...685L..67D} {685, L67}

\bibitem[\protect\citeauthoryear{{Deller} et~al.,}{{Deller}
  et~al.}{2019}]{dgb+19}
{Deller} A.~T.,  et~al., 2019, \mn@doi [\apj] {10.3847/1538-4357/ab11c7}, \href
  {http://adsabs.harvard.edu/abs/2019ApJ...875..100D} {875, 100}

\bibitem[\protect\citeauthoryear{{Deneva} et~al.,}{{Deneva}
  et~al.}{2012}]{dfc+12}
{Deneva} J.~S.,  et~al., 2012, \mn@doi [\apj] {10.1088/0004-637X/757/1/89},
  \href {http://adsabs.harvard.edu/abs/2012ApJ...757...89D} {757, 89}

\bibitem[\protect\citeauthoryear{{Desvignes} et~al.,}{{Desvignes}
  et~al.}{2016}]{dcl+16}
{Desvignes} G.,  et~al., 2016, \mn@doi [\mnras] {10.1093/mnras/stw483}, \href
  {http://adsabs.harvard.edu/abs/2016MNRAS.458.3341D} {458, 3341}

\bibitem[\protect\citeauthoryear{Detweiler}{Detweiler}{1979}]{det79}
Detweiler S.,  1979, ApJ, 234, 1100

\bibitem[\protect\citeauthoryear{{Du}, {Yang}, {Campbell}, {Janssen},
  {Stappers}  \& {Chen}}{{Du} et~al.}{2014}]{dyc+14}
{Du} Y.,  {Yang} J.,  {Campbell} R.~M.,  {Janssen} G.,  {Stappers} B.,   {Chen}
  D.,  2014, \mn@doi [\apjl] {10.1088/2041-8205/782/2/L38}, \href
  {https://ui.adsabs.harvard.edu/abs/2014ApJ...782L..38D} {782, L38}

\bibitem[\protect\citeauthoryear{Edwards \& Bailes}{Edwards \&
  Bailes}{2001}]{eb01b}
Edwards R.~T.,  Bailes M.,  2001, ApJ, 553, 801

\bibitem[\protect\citeauthoryear{{Edwards}, {Hobbs}  \& {Manchester}}{{Edwards}
  et~al.}{2006}]{ehm06}
{Edwards} R.~T.,  {Hobbs} G.~B.,   {Manchester} R.~N.,  2006, \mn@doi [\mnras]
  {10.1111/j.1365-2966.2006.10870.x}, \href
  {http://adsabs.harvard.edu/abs/2006MNRAS.372.1549E} {372, 1549}

\bibitem[\protect\citeauthoryear{{Ellis}}{{Ellis}}{2013}]{ellis13}
{Ellis} J.~A.,  2013, \mn@doi [Classical and Quantum Gravity]
  {10.1088/0264-9381/30/22/224004}, \href
  {http://adsabs.harvard.edu/abs/2013CQGra..30v4004E} {30, 224004}

\bibitem[\protect\citeauthoryear{{Faulkner} et~al.,}{{Faulkner}
  et~al.}{2004}]{fsk+04}
{Faulkner} A.~J.,  et~al., 2004, MNRAS, 355, 147

\bibitem[\protect\citeauthoryear{{Ferdman} et~al.,}{{Ferdman}
  et~al.}{2010}]{fsk+10}
{Ferdman} R.~D.,  et~al., 2010, \mn@doi [\apj] {10.1088/0004-637X/711/2/764},
  \href {http://adsabs.harvard.edu/abs/2010ApJ...711..764F} {711, 764}

\bibitem[\protect\citeauthoryear{{Folkner}, {Williams}, {Boggs}, {Park}  \&
  {Kuchynka}}{{Folkner} et~al.}{2014}]{fwb+14}
{Folkner} W.~M.,  {Williams} J.~G.,  {Boggs} D.~H.,  {Park} R.~S.,   {Kuchynka}
  P.,  2014, Interplanetary Network Progress Report, \href
  {http://adsabs.harvard.edu/abs/2014IPNPR.196C...1F} {196, 1}

\bibitem[\protect\citeauthoryear{Foster, Wolszczan  \& Camilo}{Foster
  et~al.}{1993}]{fwc93}
Foster R.~S.,  Wolszczan A.,   Camilo F.,  1993, ApJ, 410, L91

\bibitem[\protect\citeauthoryear{{Freire} \& {Wex}}{{Freire} \&
  {Wex}}{2010}]{fw10}
{Freire} P.~C.~C.,  {Wex} N.,  2010, \mn@doi [\mnras]
  {10.1111/j.1365-2966.2010.17319.x}, \href
  {http://adsabs.harvard.edu/abs/2010MNRAS.409..199F} {409, 199}

\bibitem[\protect\citeauthoryear{{Freire} et~al.,}{{Freire}
  et~al.}{2011}]{fbw+11}
{Freire} P.~C.~C.,  et~al., 2011, \mn@doi [\mnras]
  {10.1111/j.1365-2966.2010.18109.x}, \href
  {http://adsabs.harvard.edu/abs/2011MNRAS.412.2763F} {412, 2763}

\bibitem[\protect\citeauthoryear{{Freire} et~al.,}{{Freire}
  et~al.}{2012}]{fwe+12}
{Freire} P.~C.~C.,  et~al., 2012, \mn@doi [\mnras]
  {10.1111/j.1365-2966.2012.21253.x}, \href
  {http://adsabs.harvard.edu/abs/2012MNRAS.423.3328F} {423, 3328}

\bibitem[\protect\citeauthoryear{{Gaia Collaboration} et~al.,}{{Gaia
  Collaboration} et~al.}{2018}]{gaiadr2}
{Gaia Collaboration} et~al., 2018, \mn@doi [\aap]
  {10.1051/0004-6361/201833051}, \href
  {http://adsabs.harvard.edu/abs/2018A%26A...616A...1G} {616, A1}

\bibitem[\protect\citeauthoryear{{Gair}, {Romano}  \& {Taylor}}{{Gair}
  et~al.}{2015}]{grt15}
{Gair} J.~R.,  {Romano} J.~D.,   {Taylor} S.~R.,  2015, \mn@doi [\prd]
  {10.1103/PhysRevD.92.102003}, \href
  {http://adsabs.harvard.edu/abs/2015PhRvD..92j2003G} {92, 102003}

\bibitem[\protect\citeauthoryear{{Goldstein}, {Veitch}, {Sesana}  \&
  {Vecchio}}{{Goldstein} et~al.}{2018}]{gvs+18}
{Goldstein} J.~M.,  {Veitch} J.,  {Sesana} A.,   {Vecchio} A.,  2018, \mn@doi
  [\mnras] {10.1093/mnras/sty892}, \href
  {http://adsabs.harvard.edu/abs/2018MNRAS.477.5447G} {477, 5447}

\bibitem[\protect\citeauthoryear{{Guillemot} et~al.,}{{Guillemot}
  et~al.}{2016}]{gsl+16}
{Guillemot} L.,  et~al., 2016, \mn@doi [\aap] {10.1051/0004-6361/201527847},
  \href {https://ui.adsabs.harvard.edu/abs/2016A&A...587A.109G} {587, A109}

\bibitem[\protect\citeauthoryear{Guinot}{Guinot}{1988}]{gui88}
Guinot B.,  1988, A\&A, 192, 370

\bibitem[\protect\citeauthoryear{{Guo}, {Lee}  \& {Caballero}}{{Guo}
  et~al.}{2018}]{glc18}
{Guo} Y.~J.,  {Lee} K.~J.,   {Caballero} R.~N.,  2018, \mn@doi [\mnras]
  {10.1093/mnras/stx3326}, \href
  {https://ui.adsabs.harvard.edu/abs/2018MNRAS.475.3644G} {475, 3644}

\bibitem[\protect\citeauthoryear{Hellings \& Downs}{Hellings \&
  Downs}{1983}]{hd83}
Hellings R.~W.,  Downs G.~S.,  1983, ApJ, 265, L39

\bibitem[\protect\citeauthoryear{{Hessels} et~al.,}{{Hessels}
  et~al.}{2011}]{hrm+11}
{Hessels} J.~W.~T.,  et~al., 2011, in {Burgay} M.,  {D'Amico} N.,  {Esposito}
  P.,  {Pellizzoni} A.,   {Possenti} A.,  eds,  American Institute of Physics
  Conference Series Vol. 1357, American Institute of Physics Conference Series.
  pp 40--43 (\mn@eprint {arXiv} {1101.1742}), \mn@doi{10.1063/1.3615072}

\bibitem[\protect\citeauthoryear{{Hobbs} et~al.,}{{Hobbs}
  et~al.}{2004a}]{hfs+04}
{Hobbs} G.,  et~al., 2004a, MNRAS, 352, 1439

\bibitem[\protect\citeauthoryear{Hobbs, Lyne, Kramer, Martin  \& Jordan}{Hobbs
  et~al.}{2004b}]{hlk+04}
Hobbs G.,  Lyne A.~G.,  Kramer M.,  Martin C.~E.,   Jordan C.,  2004b, \mnras,
  353, 1311

\bibitem[\protect\citeauthoryear{{Hobbs}, {Edwards}  \& {Manchester}}{{Hobbs}
  et~al.}{2006}]{hem06}
{Hobbs} G.~B.,  {Edwards} R.~T.,   {Manchester} R.~N.,  2006, \mnras, 369, 655

\bibitem[\protect\citeauthoryear{{Hobbs} et~al.,}{{Hobbs}
  et~al.}{2012}]{hcm+12}
{Hobbs} G.,  et~al., 2012, \mn@doi [\mnras] {10.1111/j.1365-2966.2012.21946.x},
  \href {http://adsabs.harvard.edu/abs/2012MNRAS.427.2780H} {427, 2780}

\bibitem[\protect\citeauthoryear{{Hotan}, {Bailes}  \& {Ord}}{{Hotan}
  et~al.}{2006}]{hbo06}
{Hotan} A.~W.,  {Bailes} M.,   {Ord} S.~M.,  2006, MNRAS, 369, 1502

\bibitem[\protect\citeauthoryear{{Jacoby}}{{Jacoby}}{2005}]{jac05}
{Jacoby} B.~A.,  2005, PhD thesis, California Institute of Technology,
  California, USA

\bibitem[\protect\citeauthoryear{{Jacoby}, {Bailes}, {van Kerkwijk}, {Ord},
  {Hotan}, {Kulkarni}  \& {Anderson}}{{Jacoby} et~al.}{2003}]{jbv+03}
{Jacoby} B.~A.,  {Bailes} M.,  {van Kerkwijk} M.~H.,  {Ord} S.,  {Hotan} A.,
  {Kulkarni} S.~R.,   {Anderson} S.~B.,  2003, ApJ, \href
  {http://adsabs.harvard.edu/cgi-bin/nph-bib_query?bibcode=2003ApJ...599L..99J&amp;db_key=AST}
  {599, L99}

\bibitem[\protect\citeauthoryear{{Jacoby}, {Bailes}, {Ord}, {Knight}  \&
  {Hotan}}{{Jacoby} et~al.}{2007}]{jbo+07}
{Jacoby} B.~A.,  {Bailes} M.,  {Ord} S.~M.,  {Knight} H.~S.,   {Hotan} A.~W.,
  2007, ApJ, 656, 408

\bibitem[\protect\citeauthoryear{{Jankowski}, {van Straten}, {Keane}, {Bailes},
  {Barr}, {Johnston}  \& {Kerr}}{{Jankowski} et~al.}{2018}]{jvk+18}
{Jankowski} F.,  {van Straten} W.,  {Keane} E.~F.,  {Bailes} M.,  {Barr} E.~D.,
   {Johnston} S.,   {Kerr} M.,  2018, \mn@doi [\mnras] {10.1093/mnras/stx2476},
  \href {https://ui.adsabs.harvard.edu/abs/2018MNRAS.473.4436J} {473, 4436}

\bibitem[\protect\citeauthoryear{{Janssen}, {Stappers}, {Bassa}, {Cognard},
  {Kramer}  \& {Theureau}}{{Janssen} et~al.}{2010}]{jsb+10}
{Janssen} G.~H.,  {Stappers} B.~W.,  {Bassa} C.~G.,  {Cognard} I.,  {Kramer}
  M.,   {Theureau} G.,  2010, \mn@doi [\aap] {10.1051/0004-6361/200911728},
  \href {http://adsabs.harvard.edu/abs/2010A%26A...514A..74J} {514, A74}

\bibitem[\protect\citeauthoryear{{Jenet}, {Hobbs}, {Lee}  \&
  {Manchester}}{{Jenet} et~al.}{2005}]{jhlm05}
{Jenet} F.~A.,  {Hobbs} G.~B.,  {Lee} K.~J.,   {Manchester} R.~N.,  2005, ApJ,
  \href
  {http://adsabs.harvard.edu/cgi-bin/nph-bib_query?bibcode=2005ApJ...625L.123J&db_key=AST}
  {625, L123}

\bibitem[\protect\citeauthoryear{{Jennings}, {Kaplan}, {Chatterjee}, {Cordes}
  \& {Deller}}{{Jennings} et~al.}{2018}]{jkc+18}
{Jennings} R.~J.,  {Kaplan} D.~L.,  {Chatterjee} S.,  {Cordes} J.~M.,
  {Deller} A.~T.,  2018, \mn@doi [\apj] {10.3847/1538-4357/aad084}, \href
  {http://adsabs.harvard.edu/abs/2018ApJ...864...26J} {864, 26}

\bibitem[\protect\citeauthoryear{Johnston et~al.,}{Johnston
  et~al.}{1993}]{jlh+93}
Johnston S.,  et~al., 1993, New~Astron., 361, 613

\bibitem[\protect\citeauthoryear{{Jones} et~al.,}{{Jones}
  et~al.}{2017}]{jml+17}
{Jones} M.~L.,  et~al., 2017, \mn@doi [\apj] {10.3847/1538-4357/aa73df}, \href
  {http://adsabs.harvard.edu/abs/2017ApJ...841..125J} {841, 125}

\bibitem[\protect\citeauthoryear{Kaspi, Taylor  \& Ryba}{Kaspi
  et~al.}{1994}]{ktr94}
Kaspi V.~M.,  Taylor J.~H.,   Ryba M.,  1994, ApJ, 428, 713

\bibitem[\protect\citeauthoryear{{Keith} et~al.,}{{Keith}
  et~al.}{2013}]{kcs+13}
{Keith} M.~J.,  et~al., 2013, \mn@doi [\mnras] {10.1093/mnras/sts486}, \href
  {http://adsabs.harvard.edu/abs/2013MNRAS.429.2161K} {429, 2161}

\bibitem[\protect\citeauthoryear{{Kelley}, {Blecha}, {Hernquist}, {Sesana}  \&
  {Taylor}}{{Kelley} et~al.}{2017}]{kbh+17}
{Kelley} L.~Z.,  {Blecha} L.,  {Hernquist} L.,  {Sesana} A.,   {Taylor} S.~R.,
  2017, \mn@doi [\mnras] {10.1093/mnras/stx1638}, \href
  {http://adsabs.harvard.edu/abs/2017MNRAS.471.4508K} {471, 4508}

\bibitem[\protect\citeauthoryear{{Kelley}, {Blecha}, {Hernquist}, {Sesana}  \&
  {Taylor}}{{Kelley} et~al.}{2018}]{kbh+18}
{Kelley} L.~Z.,  {Blecha} L.,  {Hernquist} L.,  {Sesana} A.,   {Taylor} S.~R.,
  2018, \mn@doi [\mnras] {10.1093/mnras/sty689}, \href
  {http://adsabs.harvard.edu/abs/2018MNRAS.477..964K} {477, 964}

\bibitem[\protect\citeauthoryear{{Kerr} et~al.,}{{Kerr} et~al.}{2012}]{kcj+12}
{Kerr} M.,  et~al., 2012, \mn@doi [\apjl] {10.1088/2041-8205/748/1/L2}, \href
  {http://adsabs.harvard.edu/abs/2012ApJ...748L...2K} {748, L2}

\bibitem[\protect\citeauthoryear{Kramer, Xilouris, Lorimer, Doroshenko,
  Jessner, Wielebinski, Wolszczan  \& Camilo}{Kramer et~al.}{1998}]{kxl+98}
Kramer M.,  Xilouris K.~M.,  Lorimer D.~R.,  Doroshenko O.,  Jessner A.,
  Wielebinski R.,  Wolszczan A.,   Camilo F.,  1998, ApJ, 501, 270

\bibitem[\protect\citeauthoryear{{Lam} et~al.,}{{Lam} et~al.}{2017}]{lcc+17}
{Lam} M.~T.,  et~al., 2017, \mn@doi [\apj] {10.3847/1538-4357/834/1/35}, \href
  {http://adsabs.harvard.edu/abs/2017ApJ...834...35L} {834, 35}

\bibitem[\protect\citeauthoryear{Lange, Camilo, Wex, Kramer, Backer, Lyne  \&
  Doroshenko}{Lange et~al.}{2001}]{lcw+01}
Lange C.,  Camilo F.,  Wex N.,  Kramer M.,  Backer D.,  Lyne A.,   Doroshenko
  O.,  2001, MNRAS, 326, 274

\bibitem[\protect\citeauthoryear{{Lazaridis} et~al.,}{{Lazaridis}
  et~al.}{2009}]{lwj+09}
{Lazaridis} K.,  et~al., 2009, \mn@doi [\mnras]
  {10.1111/j.1365-2966.2009.15481.x}, \href
  {http://adsabs.harvard.edu/abs/2009MNRAS.400..805L} {400, 805}

\bibitem[\protect\citeauthoryear{{Lee}}{{Lee}}{2013}]{lee13}
{Lee} K.~J.,  2013, \mn@doi [Classical and Quantum Gravity]
  {10.1088/0264-9381/30/22/224016}, \href
  {http://adsabs.harvard.edu/abs/2013CQGra..30v4016L} {30, 224016}

\bibitem[\protect\citeauthoryear{{Lee}, {Jenet}  \& {Price}}{{Lee}
  et~al.}{2008}]{ljp08}
{Lee} K.~J.,  {Jenet} F.~A.,   {Price} R.~H.,  2008, \mn@doi [\apj]
  {10.1086/591080}, \href {http://adsabs.harvard.edu/abs/2008ApJ...685.1304L}
  {685, 1304}

\bibitem[\protect\citeauthoryear{{Lee}, {Jenet}, {Price}, {Wex}  \&
  {Kramer}}{{Lee} et~al.}{2010}]{ljp+10}
{Lee} K.,  {Jenet} F.~A.,  {Price} R.~H.,  {Wex} N.,   {Kramer} M.,  2010,
  \mn@doi [\apj] {10.1088/0004-637X/722/2/1589}, \href
  {https://ui.adsabs.harvard.edu/abs/2010ApJ...722.1589L} {722, 1589}

\bibitem[\protect\citeauthoryear{{Lee}, {Wex}, {Kramer}, {Stappers}, {Bassa},
  {Janssen}, {Karuppusamy}  \& {Smits}}{{Lee} et~al.}{2011}]{lwk+11}
{Lee} K.~J.,  {Wex} N.,  {Kramer} M.,  {Stappers} B.~W.,  {Bassa} C.~G.,
  {Janssen} G.~H.,  {Karuppusamy} R.,   {Smits} R.,  2011, \mn@doi [\mnras]
  {10.1111/j.1365-2966.2011.18622.x}, \href
  {https://ui.adsabs.harvard.edu/abs/2011MNRAS.414.3251L} {414, 3251}

\bibitem[\protect\citeauthoryear{{Lee} et~al.,}{{Lee} et~al.}{2014}]{lbj+14}
{Lee} K.~J.,  et~al., 2014, \mn@doi [\mnras] {10.1093/mnras/stu664}, \href
  {http://adsabs.harvard.edu/abs/2014MNRAS.441.2831L} {441, 2831}

\bibitem[\protect\citeauthoryear{{Lentati}, {Alexander}, {Hobson}, {Feroz},
  {van Haasteren}, {Lee}  \& {Shannon}}{{Lentati} et~al.}{2014}]{lah+14}
{Lentati} L.,  {Alexander} P.,  {Hobson} M.~P.,  {Feroz} F.,  {van Haasteren}
  R.,  {Lee} K.~J.,   {Shannon} R.~M.,  2014, \mn@doi [\mnras]
  {10.1093/mnras/stt2122}, \href
  {http://adsabs.harvard.edu/abs/2014MNRAS.437.3004L} {437, 3004}

\bibitem[\protect\citeauthoryear{{Lentati} et~al.,}{{Lentati}
  et~al.}{2015}]{ltm+15}
{Lentati} L.,  et~al., 2015, \mn@doi [\mnras] {10.1093/mnras/stv1538}, \href
  {http://adsabs.harvard.edu/abs/2015MNRAS.453.2576L} {453, 2576}

\bibitem[\protect\citeauthoryear{{Lentati} et~al.,}{{Lentati}
  et~al.}{2016}]{lsc+16}
{Lentati} L.,  et~al., 2016, \mn@doi [\mnras] {10.1093/mnras/stw395}, \href
  {http://adsabs.harvard.edu/abs/2016MNRAS.458.2161L} {458, 2161}

\bibitem[\protect\citeauthoryear{{Levin} et~al.,}{{Levin}
  et~al.}{2016}]{lmj+16}
{Levin} L.,  et~al., 2016, \mn@doi [\apj] {10.3847/0004-637X/818/2/166}, \href
  {http://adsabs.harvard.edu/abs/2016ApJ...818..166L} {818, 166}

\bibitem[\protect\citeauthoryear{{Lommen}, {Zepka}, {Backer}, {McLaughlin},
  {Cordes}, {Arzoumanian}  \& {Xilouris}}{{Lommen} et~al.}{2000}]{lzb+00}
{Lommen} A.~N.,  {Zepka} A.,  {Backer} D.~C.,  {McLaughlin} M.,  {Cordes}
  J.~M.,  {Arzoumanian} Z.,   {Xilouris} K.,  2000, \apj, 545, 1007

\bibitem[\protect\citeauthoryear{Lorimer \& Kramer}{Lorimer \&
  Kramer}{2005}]{lk05}
Lorimer D.~R.,  Kramer M.,  2005, {Handbook of Pulsar Astronomy}.
Cambridge University Press

\bibitem[\protect\citeauthoryear{Lorimer et~al.,}{Lorimer
  et~al.}{1995}]{lnl+95}
Lorimer D.~R.,  et~al., 1995, ApJ, 439, 933

\bibitem[\protect\citeauthoryear{Lorimer, Lyne, Bailes, Manchester, D'Amico,
  Stappers, Johnston  \& Camilo}{Lorimer et~al.}{1996}]{llb+96}
Lorimer D.~R.,  Lyne A.~G.,  Bailes M.,  Manchester R.~N.,  D'Amico N.,
  Stappers B.~W.,  Johnston S.,   Camilo F.,  1996, MNRAS, 283, 1383

\bibitem[\protect\citeauthoryear{{Lorimer} et~al.,}{{Lorimer}
  et~al.}{2006}]{lfl+06}
{Lorimer} D.~R.,  et~al., 2006, MNRAS, 372, 777

\bibitem[\protect\citeauthoryear{Lundgren, Zepka  \& Cordes}{Lundgren
  et~al.}{1995}]{lzc95}
Lundgren S.~C.,  Zepka A.~F.,   Cordes J.~M.,  1995, ApJ, 453, 419

\bibitem[\protect\citeauthoryear{{Lynch} et~al.,}{{Lynch}
  et~al.}{2013}]{lbr+13}
{Lynch} R.~S.,  et~al., 2013, \mn@doi [\apj] {10.1088/0004-637X/763/2/81},
  \href {http://adsabs.harvard.edu/abs/2013ApJ...763...81L} {763, 81}

\bibitem[\protect\citeauthoryear{Lyne, Brinklow, Middleditch, Kulkarni, Backer
  \& Clifton}{Lyne et~al.}{1987}]{lbm+87}
Lyne A.~G.,  Brinklow A.,  Middleditch J.,  Kulkarni S.~R.,  Backer D.~C.,
  Clifton T.~R.,  1987, New~Astron., 328, 399

\bibitem[\protect\citeauthoryear{{Madison}, {Chernoff}  \& {Cordes}}{{Madison}
  et~al.}{2017}]{mcc17}
{Madison} D.~R.,  {Chernoff} D.~F.,   {Cordes} J.~M.,  2017, \mn@doi [\prd]
  {10.1103/PhysRevD.96.123016}, \href
  {http://adsabs.harvard.edu/abs/2017PhRvD..96l3016M} {96, 123016}

\bibitem[\protect\citeauthoryear{{Madison} et~al.,}{{Madison}
  et~al.}{2019}]{mca+19}
{Madison} D.~R.,  et~al., 2019, \mn@doi [\apj] {10.3847/1538-4357/ab01fd},
  \href {https://ui.adsabs.harvard.edu/abs/2019ApJ...872..150M} {872, 150}

\bibitem[\protect\citeauthoryear{{Manchester} et~al.,}{{Manchester}
  et~al.}{2013}]{mhb+13}
{Manchester} R.~N.,  et~al., 2013, \mn@doi [\pasa] {10.1017/pasa.2012.017},
  \href {http://adsabs.harvard.edu/abs/2013PASA...30...17M} {30, e017}

\bibitem[\protect\citeauthoryear{{Mingarelli}, {Grover}, {Sidery}, {Smith}  \&
  {Vecchio}}{{Mingarelli} et~al.}{2012}]{mgs+12}
{Mingarelli} C.~M.~F.,  {Grover} K.,  {Sidery} T.,  {Smith} R.~J.~E.,
  {Vecchio} A.,  2012, \mn@doi [Physical Review Letters]
  {10.1103/PhysRevLett.109.081104}, \href
  {http://adsabs.harvard.edu/abs/2012PhRvL.109h1104M} {109, 081104}

\bibitem[\protect\citeauthoryear{{Mingarelli} et~al.,}{{Mingarelli}
  et~al.}{2017}]{mls+17}
{Mingarelli} C.~M.~F.,  et~al., 2017, \mn@doi [Nature Astronomy]
  {doi:10.1038/s41550-017-0299-6}, 1, 886

\bibitem[\protect\citeauthoryear{{Mingarelli}, {Anderson}, {Bedell}  \&
  {Spergel}}{{Mingarelli} et~al.}{2018}]{mab+19}
{Mingarelli} C.~M.~F.,  {Anderson} L.,  {Bedell} M.,   {Spergel} D.~N.,  2018,
  arXiv e-prints, \href {http://adsabs.harvard.edu/abs/2018arXiv181206262M} {1812.06262}

\bibitem[\protect\citeauthoryear{Navarro, de Bruyn, Frail, Kulkarni  \&
  Lyne}{Navarro et~al.}{1995}]{nbf+95}
Navarro J.,  de Bruyn G.,  Frail D.,  Kulkarni S.~R.,   Lyne A.~G.,  1995, ApJ,
  455, L55

\bibitem[\protect\citeauthoryear{Nicastro, Lyne, Lorimer, Harrison, Bailes  \&
  Skidmore}{Nicastro et~al.}{1995}]{nll+95}
Nicastro L.,  Lyne A.~G.,  Lorimer D.~R.,  Harrison P.~A.,  Bailes M.,
  Skidmore B.~D.,  1995, MNRAS, 273, L68

\bibitem[\protect\citeauthoryear{Nice, Taylor  \& Fruchter}{Nice
  et~al.}{1993}]{ntf93}
Nice D.~J.,  Taylor J.~H.,   Fruchter A.~S.,  1993, ApJ, 402, L49

\bibitem[\protect\citeauthoryear{Nice, Splaver  \& Stairs}{Nice
  et~al.}{2001}]{nss01}
Nice D.~J.,  Splaver E.~M.,   Stairs I.~H.,  2001, ApJ, 549, 516

\bibitem[\protect\citeauthoryear{{Niu}, {Hobbs}, {Wang}  \& {Dai}}{{Niu}
  et~al.}{2017}]{nhw+17}
{Niu} Z.-X.,  {Hobbs} G.,  {Wang} J.-B.,   {Dai} S.,  2017, \mn@doi [Research
  in Astronomy and Astrophysics] {10.1088/1674-4527/17/10/103}, \href
  {https://ui.adsabs.harvard.edu/abs/2017RAA....17..103N} {17, 103}

\bibitem[\protect\citeauthoryear{{O'Beirne}, {Cornish}, {Vigeland}  \&
  {Taylor}}{{O'Beirne} et~al.}{2019}]{ocv+19}
{O'Beirne} L.,  {Cornish} N.~J.,  {Vigeland} S.~J.,   {Taylor} S.~R.,  2019,
  arXiv e-prints, \href {http://adsabs.harvard.edu/abs/2019arXiv190402744O} {1904.02744}

\bibitem[\protect\citeauthoryear{{Os{\l}owski}, {van Straten}, {Hobbs},
  {Bailes}  \& {Demorest}}{{Os{\l}owski} et~al.}{2011}]{ovh+11}
{Os{\l}owski} S.,  {van Straten} W.,  {Hobbs} G.~B.,  {Bailes} M.,   {Demorest}
  P.,  2011, \mn@doi [\mnras] {10.1111/j.1365-2966.2011.19578.x}, \href
  {http://adsabs.harvard.edu/abs/2011MNRAS.418.1258O} {418, 1258}

\bibitem[\protect\citeauthoryear{{Os{\l}owski}, {van Straten}, {Demorest}  \&
  {Bailes}}{{Os{\l}owski} et~al.}{2013}]{ovd+13}
{Os{\l}owski} S.,  {van Straten} W.,  {Demorest} P.,   {Bailes} M.,  2013,
  \mn@doi [\mnras] {10.1093/mnras/sts662}, \href
  {http://adsabs.harvard.edu/abs/2013MNRAS.430..416O} {430, 416}

\bibitem[\protect\citeauthoryear{{Perera} et~al.,}{{Perera}
  et~al.}{2018}]{psb+18}
{Perera} B.~B.~P.,  et~al., 2018, \mn@doi [\mnras] {10.1093/mnras/sty1116},
  \href {http://adsabs.harvard.edu/abs/2018MNRAS.478..218P} {478, 218}

\bibitem[\protect\citeauthoryear{Petit}{Petit}{2003}]{pet03b}
Petit G.,  2003, in 35th Annual Precise Time and Time Interval (PTTI) Meeting,
  San Diego, December 2003. pp 307--317

\bibitem[\protect\citeauthoryear{{Porayko} et~al.,}{{Porayko}
  et~al.}{2018}]{pzl+18}
{Porayko} N.~K.,  et~al., 2018, \mn@doi [\prd] {10.1103/PhysRevD.98.102002},
  \href {http://adsabs.harvard.edu/abs/2018PhRvD..98j2002P} {98, 102002}

\bibitem[\protect\citeauthoryear{Radhakrishnan \& Srinivasan}{Radhakrishnan \&
  Srinivasan}{1982}]{rs82}
Radhakrishnan V.,  Srinivasan G.,  1982, \cursci, 51, 1096

\bibitem[\protect\citeauthoryear{{Ransom} et~al.,}{{Ransom}
  et~al.}{2011}]{rrc+11}
{Ransom} S.~M.,  et~al., 2011, \mn@doi [\apjl] {10.1088/2041-8205/727/1/L16},
  \href {http://adsabs.harvard.edu/abs/2011ApJ...727L..16R} {727, L16}

\bibitem[\protect\citeauthoryear{Ray, Thorsett, Jenet, van Kerkwijk, Kulkarni,
  Prince, Sandhu  \& Nice}{Ray et~al.}{1996}]{rtj+96}
Ray P.~S.,  Thorsett S.~E.,  Jenet F.~A.,  van Kerkwijk M.~H.,  Kulkarni S.~R.,
   Prince T.~A.,  Sandhu J.~S.,   Nice D.~J.,  1996, ApJ, 470, 1103

\bibitem[\protect\citeauthoryear{{Reardon} et~al.,}{{Reardon}
  et~al.}{2016}]{rhc+16}
{Reardon} D.~J.,  et~al., 2016, \mn@doi [\mnras] {10.1093/mnras/stv2395}, \href
  {http://adsabs.harvard.edu/abs/2016MNRAS.455.1751R} {455, 1751}

\bibitem[\protect\citeauthoryear{{Rosado}, {Sesana}  \& {Gair}}{{Rosado}
  et~al.}{2015}]{rsg15}
{Rosado} P.~A.,  {Sesana} A.,   {Gair} J.,  2015, \mn@doi [\mnras]
  {10.1093/mnras/stv1098}, \href
  {http://adsabs.harvard.edu/abs/2015MNRAS.451.2417R} {451, 2417}

\bibitem[\protect\citeauthoryear{Segelstein, Rawley, Stinebring, Fruchter  \&
  Taylor}{Segelstein et~al.}{1986}]{srs+86}
Segelstein D.~J.,  Rawley L.~A.,  Stinebring D.~R.,  Fruchter A.~S.,   Taylor
  J.~H.,  1986, New~Astron., 322, 714

\bibitem[\protect\citeauthoryear{{Sesana} \& {Vecchio}}{{Sesana} \&
  {Vecchio}}{2010}]{sv10}
{Sesana} A.,  {Vecchio} A.,  2010, \mn@doi [\prd] {10.1103/PhysRevD.81.104008},
  \href {http://adsabs.harvard.edu/abs/2010PhRvD..81j4008S} {81, 104008}

\bibitem[\protect\citeauthoryear{{Shannon} et~al.,}{{Shannon}
  et~al.}{2015}]{srl+15}
{Shannon} R.~M.,  et~al., 2015, \mn@doi [Science] {10.1126/science.aab1910},
  \href {http://adsabs.harvard.edu/abs/2015Sci...349.1522S} {349, 1522}

\bibitem[\protect\citeauthoryear{{Siemens}, {Ellis}, {Jenet}  \&
  {Romano}}{{Siemens} et~al.}{2013}]{sej+13}
{Siemens} X.,  {Ellis} J.,  {Jenet} F.,   {Romano} J.~D.,  2013, \mn@doi
  [Classical and Quantum Gravity] {10.1088/0264-9381/30/22/224015}, \href
  {http://adsabs.harvard.edu/abs/2013CQGra..30v4015S} {30, 224015}

\bibitem[\protect\citeauthoryear{Stairs et~al.,}{Stairs et~al.}{2005}]{sfl+05}
Stairs I.~H.,  et~al., 2005, ApJ, 632, 1060

\bibitem[\protect\citeauthoryear{{Stovall} et~al.,}{{Stovall}
  et~al.}{2014}]{slr+14}
{Stovall} K.,  et~al., 2014, \mn@doi [\apj] {10.1088/0004-637X/791/1/67}, \href
  {http://adsabs.harvard.edu/abs/2014ApJ...791...67S} {791, 67}

\bibitem[\protect\citeauthoryear{{Taylor}, {Ellis}  \& {Gair}}{{Taylor}
  et~al.}{2014}]{teg14}
{Taylor} S.,  {Ellis} J.,   {Gair} J.,  2014, \mn@doi [\prd]
  {10.1103/PhysRevD.90.104028}, \href
  {https://ui.adsabs.harvard.edu/abs/2014PhRvD..90j4028T} {90, 104028}

\bibitem[\protect\citeauthoryear{{Taylor}, {Vallisneri}, {Ellis}, {Mingarelli},
  {Lazio}  \& {van Haasteren}}{{Taylor} et~al.}{2016}]{tve+16}
{Taylor} S.~R.,  {Vallisneri} M.,  {Ellis} J.~A.,  {Mingarelli} C.~M.~F.,
  {Lazio} T.~J.~W.,   {van Haasteren} R.,  2016, \mn@doi [\apjl]
  {10.3847/2041-8205/819/1/L6}, \href
  {http://adsabs.harvard.edu/abs/2016ApJ...819L...6T} {819, L6}

\bibitem[\protect\citeauthoryear{{Tiburzi} et~al.,}{{Tiburzi}
  et~al.}{2019}]{tvs+19}
{Tiburzi} C.,  et~al., 2019, arXiv e-prints, \href
  {https://ui.adsabs.harvard.edu/abs/2019arXiv190502989T} {p. arXiv:1905.02989}

\bibitem[\protect\citeauthoryear{Toscano, Bailes, Manchester  \&
  Sandhu}{Toscano et~al.}{1998}]{tbms98}
Toscano M.,  Bailes M.,  Manchester R.,   Sandhu J.,  1998, ApJ, 506, 863

\bibitem[\protect\citeauthoryear{{Verbiest} et~al.,}{{Verbiest}
  et~al.}{2008}]{vbv+08}
{Verbiest} J.~P.~W.,  et~al., 2008, \mn@doi [\apj] {10.1086/529576}, \href
  {http://adsabs.harvard.edu/abs/2008ApJ...679..675V} {679, 675}

\bibitem[\protect\citeauthoryear{{Verbiest} et~al.,}{{Verbiest}
  et~al.}{2009}]{vbc+09}
{Verbiest} J.~P.~W.,  et~al., 2009, \mn@doi [\mnras]
  {10.1111/j.1365-2966.2009.15508.x}, \href
  {https://ui.adsabs.harvard.edu/abs/2009MNRAS.400..951V} {400, 951}

\bibitem[\protect\citeauthoryear{{Verbiest}, {Weisberg}, {Chael}, {Lee}  \&
  {Lorimer}}{{Verbiest} et~al.}{2012}]{vwc+12}
{Verbiest} J.~P.~W.,  {Weisberg} J.~M.,  {Chael} A.~A.,  {Lee} K.~J.,
  {Lorimer} D.~R.,  2012, \mn@doi [\apj] {10.1088/0004-637X/755/1/39}, \href
  {http://adsabs.harvard.edu/abs/2012ApJ...755...39V} {755, 39}

\bibitem[\protect\citeauthoryear{{Verbiest} et~al.,}{{Verbiest}
  et~al.}{2016}]{vlh+16}
{Verbiest} J.~P.~W.,  et~al., 2016, \mn@doi [\mnras] {10.1093/mnras/stw347},
  \href {http://adsabs.harvard.edu/abs/2016MNRAS.458.1267V} {458, 1267}

\bibitem[\protect\citeauthoryear{{Wang} et~al.,}{{Wang} et~al.}{2015}]{whc+15}
{Wang} J.~B.,  et~al., 2015, \mn@doi [\mnras] {10.1093/mnras/stu2137}, \href
  {http://adsabs.harvard.edu/abs/2015MNRAS.446.1657W} {446, 1657}

\bibitem[\protect\citeauthoryear{Wex}{Wex}{1999}]{wex99}
Wex N.,  1999, Private communication (Implementation of Laplace-Lagrange
  parameters in TEMPO)

\bibitem[\protect\citeauthoryear{{Yao}, {Manchester}  \& {Wang}}{{Yao}
  et~al.}{2017}]{ymw17}
{Yao} J.~M.,  {Manchester} R.~N.,   {Wang} N.,  2017, \mn@doi [\apj]
  {10.3847/1538-4357/835/1/29}, \href
  {http://adsabs.harvard.edu/abs/2017ApJ...835...29Y} {835, 29}

\bibitem[\protect\citeauthoryear{{Yardley} et~al.,}{{Yardley}
  et~al.}{2010}]{yhj+10}
{Yardley} D.~R.~B.,  et~al., 2010, \mn@doi [\mnras]
  {10.1111/j.1365-2966.2010.16949.x}, \href
  {http://adsabs.harvard.edu/abs/2010MNRAS.407..669Y} {407, 669}

\bibitem[\protect\citeauthoryear{{Zhu} et~al.,}{{Zhu} et~al.}{2014}]{zhw+14}
{Zhu} X.-J.,  et~al., 2014, \mn@doi [\mnras] {10.1093/mnras/stu1717}, \href
  {http://adsabs.harvard.edu/abs/2014MNRAS.444.3709Z} {444, 3709}

\bibitem[\protect\citeauthoryear{{Zhu} et~al.,}{{Zhu} et~al.}{2015}]{zsd+15}
{Zhu} W.~W.,  et~al., 2015, \mn@doi [\apj] {10.1088/0004-637X/809/1/41}, \href
  {http://adsabs.harvard.edu/abs/2015ApJ...809...41Z} {809, 41}

\bibitem[\protect\citeauthoryear{{Zhu}, {Wen}, {Xiong}, {Xu}, {Wang},
  {Mohanty}, {Hobbs}  \& {Manchester}}{{Zhu} et~al.}{2016}]{zwx+16}
{Zhu} X.-J.,  {Wen} L.,  {Xiong} J.,  {Xu} Y.,  {Wang} Y.,  {Mohanty} S.~D.,
  {Hobbs} G.,   {Manchester} R.~N.,  2016, \mn@doi [\mnras]
  {10.1093/mnras/stw1446}, \href
  {http://adsabs.harvard.edu/abs/2016MNRAS.461.1317Z} {461, 1317}

\bibitem[\protect\citeauthoryear{{van Haasteren} \& {Levin}}{{van Haasteren} \&
  {Levin}}{2010}]{vHL10}
{van Haasteren} R.,  {Levin} Y.,  2010, \mn@doi [\mnras]
  {10.1111/j.1365-2966.2009.15885.x}, \href
  {http://adsabs.harvard.edu/abs/2010MNRAS.401.2372V} {401, 2372}

\makeatother
\end{thebibliography}
\bibliographystyle{mnras}

\appendix
\section{Timing models}
\label{app}

In this appendix, we present the updated timing solutions for all 65 MSPs according to VersionB as described in \S~\ref{verB} based on this most up-to-date IPTA data combination.

\begin{table*}
\begin{center}
\caption{
Timing solutions of pulsars based on VersionB described in \S~\ref{verB}. The values in parentheses represent the 1-$\sigma$ uncertainty of the last displayed digit for the parameter. The description of parameters is given in \S~\ref{procedure}. 
}
\label{timing}
\begin{tabular}{lccccccc}
\hline
\multicolumn{1}{l}{} &
\multicolumn{1}{c}{J0023+0923} &
\multicolumn{1}{c}{J0030+0451} &
\multicolumn{1}{c}{J0034-0534} &
\multicolumn{1}{c}{J0218+4232} \\
\hline
MJD range &   $   55757  -   56600  $ & $   51275  -   56780  $ & $   51769  -   56707  $ & $   50370  -   56788  $ \\
Number of ToAs &   $     4373$ & $     3362$ & $      276$ &
$     1196$   \\
Weighted rms timing residual ($\mu$s) &   $   1.34$ & $   3.32$ & $   4.27$ & $   7.82$ \\
Reduced $\chi^2$  &   $   1.01$ & $   1.01$ & $   0.93$ & $   1.02$ \\
Reference epoch &   $  56200$ & $  55000$ & $  55000$ & $  55000$ \\
Units & TCB & TCB & TCB & TCB \smallskip \\
Right ascension, RA (J2000)   & $00$:$23$:$                      16.87910(    2) $ & $00$:$30$:$                      27.42838(    3) $ & $00$:$34$:$                      21.83424(    8) $ & $02$:$18$:$                       06.35731(    2) $ \\
\hline
\end{tabular}
\end{center}
\end{table*}

\begin{table*}
\begin{center}
\contcaption{}
\resizebox{\textwidth}{!}{%
\begin{tabular}{lccccccc}
\hline
\multicolumn{1}{l}{} &
\multicolumn{1}{c}{J0621$+$1002} &
\multicolumn{1}{c}{J0645$+$5158} &
\multicolumn{1}{c}{J0711$-$6830} &
\multicolumn{1}{c}{J0751$+$1807} \\
\hline
MJD range &   $   52481  -   56782  $ & $   55703  -   56587  $ & $   49373  -   55620  $ & $   50363  -   56794  $ \\
Number of ToAs &   $      682$ & $     2891$ & $      507$ & 
$     1491$   \\
Weighted rms timing residual ($\mu$s) &   $   9.52$ & $   0.92$ & $   1.44$ & $   3.01$ \\
Reduced $\chi^2$  &   $   1.01$ & $   1.00$ & $   1.00$ & $   1.01$ \\
Reference epoch &   $  55000$ & $  56143$ & $  55000$ & $  55000$ \\
Units & TCB & TCB & TCB & TCB \smallskip \\
Right ascension, RA (J2000)   & $06$:$21$:$                      22.11438(    2) $ & $06$:$45$:$                     59.081909(    4) $ & $07$:$11$:$                      54.18529(    1) $ & $07$:$51$:$                       09.15535(    2) $ \\
\hline
\end{tabular}}
\begin{tabular}{l}
$^\dagger$ Ecliptic coordinates are used for PSR J1022$+$1001 and thus, ecliptic longitude $\lambda$ and latitude $\beta$, and their proper motions are constrained.
\end{tabular}
\end{center}
\end{table*}

\begin{table*}
\begin{center}
\contcaption{}
\begin{tabular}{lccccccc}
\hline
\multicolumn{1}{l}{} &
\multicolumn{1}{c}{J1024-0719} &
\multicolumn{1}{c}{J1045-4509} &
\multicolumn{1}{c}{J1455-3330} &
\multicolumn{1}{c}{J1600-3053} \\
\hline
MJD range &   $   50117  -   56766  $ & $   49406  -   55620  $ & $   53217  -   56752  $ & $   52301  -   56796  $ \\
Number of ToAs &   $     5865$ & $      605$ & $     5507$ & $     9006$   \\
Weighted rms timing residual ($\mu$s) &   $   7.29$ & $   3.19$ & $   4.13$ & $   0.95$ \\
Reduced $\chi^2$  &   $   0.98$ & $   1.00$ & $   1.01$ & $   1.00$ \\
Reference epoch &   $  55000$ & $  55000$ & $  55000$ & $  55000$ \\
Units & TCB & TCB & TCB & TCB \smallskip \\
Right ascension, RA (J2000)  & $10$:$24$:$                     38.675401(    3) $ & $10$:$45$:$                      50.18621(    2) $ & $14$:$55$:$                     47.969867(    9) $ & $16$:$00$:$                     51.903355(    3) $ \\
\hline
\end{tabular}
\end{center}
\end{table*}

\begin{table*}
\begin{center}
\contcaption{}
\begin{tabular}{lccccccc}
\hline
\multicolumn{1}{l}{} &
\multicolumn{1}{c}{J1713+0747} &
\multicolumn{1}{c}{J1721-2457} &
\multicolumn{1}{c}{J1730-2304} &
\multicolumn{1}{c}{J1732-5049} \\
\hline
MJD range &   $   48849  -   57053  $ & $   52076  -   56737  $ & $   49422  -   56831  $ & $   52646  -   55583  $ \\
Number of ToAs &   $    17487$ & $      150$ & $      646$ & 
$      242$   \\
Weighted rms timing residual ($\mu$s) &   $   0.24$ & $  12.21$ & $   1.67$ & $   2.80$ \\
Reduced $\chi^2$  &   $   0.99$ & $   0.98$ & $   0.92$ & $   0.92$ \\
Reference epoch &   $  55000$ & $  55000$ & $  55000$ & $  55000$ \\
Units & TCB & TCB & TCB & TCB \smallskip \\
Right ascension, RA (J2000)   & $17$:$13$:$                    49.5331960(    4) $ & $17$:$21$:$                        05.4980(    2) $ & $17$:$30$:$                      21.66836(    7) $ & $17$:$32$:$                      47.76668(    2) $ \\
\hline
\end{tabular}
\end{center}
\end{table*}

\begin{table*}
\begin{center}
\contcaption{}
\begin{tabular}{lccccccc}
\hline
\multicolumn{1}{l}{} &
\multicolumn{1}{c}{J1751$-$2857} &
\multicolumn{1}{c}{J1801$-$1417} &
\multicolumn{1}{c}{J1802$-$2124} &
\multicolumn{1}{c}{J1804$-$2717} \\
\hline
MJD range &   $   53746  -   56782  $ & $   54206  -   56782  $ & $   54187  -   56832  $ & $   53766  -   56828  $ \\
Number of ToAs &   $      144$ & $      126$ & $      522$ & $      116$   \\
Weighted rms timing residual ($\mu$s) &   $   3.12$ & $   2.76$ & $   2.76$ & $   3.72$ \\
Reduced $\chi^2$  &   $   0.92$ & $   0.90$ & $   1.02$ & $   0.98$ \\
Reference epoch &   $  55000$ & $  55000$ & $  55000$ & $  55000$ \\
Units & TCB & TCB & TCB & TCB \smallskip \\
Right ascension, RA (J2000)  & $17$:$51$:$                      32.69320(    2) $ & $18$:$01$:$                      51.07335(    2) $ & $18$:$02$:$                       05.33524(    2) $ & $18$:$04$:$                      21.13311(    2) $ \\
\hline
\end{tabular}
\end{center}
\end{table*}

\begin{table*}
\begin{center}
\contcaption{}
\begin{tabular}{lccccccc}
\hline
\multicolumn{1}{l}{} &
\multicolumn{1}{c}{J1857$+$0943} &
\multicolumn{1}{c}{J1903$+$0327} &
\multicolumn{1}{c}{J1909$-$3744} &
\multicolumn{1}{c}{J1910$+$1256} \\
\hline
MJD range &   $   46401  -   56782  $ & $   55135  -   56593  $ & $   53040  -   56993  $ & $   53370  -   56829  $ \\
Number of ToAs &   $     5004$ & $     1802$ & $    11483$ & $     2743$   \\
Weighted rms timing residual ($\mu$s) &   $   1.25$ & $   2.85$ & $   0.20$ & $   1.42$ \\
Reduced $\chi^2$  &   $   0.99$ & $   0.98$ & $   1.00$ & $   1.01$ \\
Reference epoch &   $  55000$ & $  55712$ & $  55000$ & $  55000$ \\
Units & TCB & TCB & TCB & TCB \smallskip \\
Right ascension, RA (J2000)  & $18$:$57$:$                     36.390622(    3) $ & $19$:$03$:$                       05.79288(    2) $ & $19$:$09$:$                    47.4335840(    4) $ & $19$:$10$:$                      09.701469(    6) $ \\
\hline
\end{tabular}
\end{center}
\end{table*}

\begin{table*}
\begin{center}
\contcaption{}
\begin{tabular}{lccccccc}
\hline
\multicolumn{1}{l}{} &
\multicolumn{1}{c}{J1939+2134} &
\multicolumn{1}{c}{J1944+0907} &
\multicolumn{1}{c}{J1949+3106} &
\multicolumn{1}{c}{J1955+2908} \\
\hline
MJD range &   $   46024  -   56779  $ & $   54505  -   56593  $ & $   56138  -   56595  $ & $   53812  -   56782  $ \\
Number of ToAs &   $    13659$ & $     1696$ & $     1409$ & 
$     1459$   \\
Weighted rms timing residual ($\mu$s) &   $  50.63$ & $   2.23$ & $   4.61$ & $   3.62$ \\
Reduced $\chi^2$  &   $   0.94$ & $   0.99$ & $   1.00$ & $   1.00$ \\
Reference epoch &   $  55000$ & $  55000$ & $  56367$ & $  55000$ \\
Units & TCB & TCB & TCB & TCB \smallskip \\
Right ascension, RA (J2000)  & $19$:$39$:$                     38.561249(    2) $ & $19$:$44$:$                       09.32685(    2) $ & $19$:$49$:$                       29.6373(    2) $ & $19$:$55$:$                      27.87578(    1) $ \\
\hline
\end{tabular}
\end{center}
\end{table*}

\begin{table*}
\begin{center}
\contcaption{}
\begin{tabular}{lccccccc}
\hline
\multicolumn{1}{l}{} &
\multicolumn{1}{c}{J2043$+$1711} &
\multicolumn{1}{c}{J2124$-$3358} &
\multicolumn{1}{c}{J2129$-$5721} &
\multicolumn{1}{c}{J2145$-$0750} \\
\hline
MJD range &   $   55757  -   56593  $ & $   49489  -   56796  $ & $   49987  -   55618  $ & $   49001  -   56762  $ \\
Number of ToAs &   $     1382$ & $     1182$ & $      373$ & $     8456$   \\
Weighted rms timing residual ($\mu$s) &   $   0.63$ & $   3.00$ & $   1.03$ & $   6.08$ \\
Reduced $\chi^2$  &   $   0.99$ & $   0.97$ & $   1.04$ & $   1.00$ \\
Reference epoch &   $  56175$ & $  55000$ & $  55000$ & $  55000$ \\
Units & TCB & TCB & TCB & TCB \smallskip \\
Right ascension, RA (J2000)  & $20$:$43$:$                     20.882167(    2) $ & $21$:$24$:$                      43.84783(    1) $ & $21$:$29$:$                     22.768535(    8) $ & $21$:$45$:$                     50.460606(    7) $ \\
\hline
\end{tabular}
\end{center}
\end{table*}

\begin{table*}
\begin{center}
\contcaption{}
\begin{tabular}{lc}
\hline
\multicolumn{1}{l}{} &
\multicolumn{1}{c}{J2322$+$2057} \\
\hline
MJD range &   $   53905  -   56789  $ \\
Number of ToAs &   $      229$ \\
Weighted rms timing residual ($\mu$s) &   $   6.74$ \\
Reduced $\chi^2$  &   $   0.97$ \\
Reference epoch &   $  55000$ \\
Units & TCB \smallskip \\
Right ascension, RA (J2000)  & $23$:$22$:$                      22.33519(    7) $ \\
Declination, DEC (J2000)  & $+20$:$57$:$                         02.677(    1) $ \\
Proper motion in RA (mas~yr$^{-1}$)   & $                         -18.5(    3) $ \\
Proper motion in DEC (mas~yr$^{-1}$)   & $                         -15.3(    5) $ \\
Spin frequency, $f$ (s$^{-1}$)   & $               207.96816335834(    7) $ \\
$\dot{f}$ (s$^{-2}$)   & $        -4.181(    9) \times10^{-16} $ \\
Dispersion measure, ${\rm DM}$ (cm$^{-3}$~pc)   & $                         13.35(    3) $ \\
$\dot{\rm DM}$ (cm$^{-3}$~pc~yr$^{-1}$)   & $                        -0.001(    5) $ \\
$\ddot{\rm DM}$ (cm$^{-3}$~pc~yr$^{-2}$)   & $                        -0.001(    1) $ \\
\hline
\end{tabular}
\end{center}
\end{table*}

\end{document}